\documentclass[fleqn,usenatbib,useAMS]{mnras}
\usepackage{deluxetable}
\usepackage{epic,eepic}
\usepackage{graphicx}
\usepackage{color}
\usepackage{epstopdf}
\usepackage{epsfig}
\usepackage{wasysym}

\usepackage{amsmath, amssymb}
\usepackage{orcidlink}

\usepackage[english]{babel}
\usepackage[latin1, applemac]{inputenc}

\DeclareMathOperator{\sinc}{sinc}

\newcommand{\Rom}[1]
    {\MakeUppercase{\romannumeral #1}}

\title[Flares in the changing look AGN Mrk 590.]{Flares in the changing look AGN Mrk 590. \Rom{1}: The UV response to X-ray outbursts suggests a more complex reprocessing geometry than a standard disk.}

\author[Lawther et al.]{D. Lawther\orcidlink{0000-0003-3959-5534},$^1$ M. Vestergaard\orcidlink{0000-0001-9191-9837},$^{1,2}$ S. Raimundo\orcidlink{0000-0002-6248-398X},$^{2,3,4}$ J.Y. Koay\orcidlink{0000-0002-7029-6658},$^{5}$ B.M. Peterson\orcidlink{0000-0001-6481-5397},$^6$ \newauthor X. Fan\orcidlink{0000-0003-3310-0131},$^{1}$ D. Grupe\orcidlink{0000-0002-9961-3661},$^7$  S. Mathur\orcidlink{0000-0002-4822-3559}$^8$\\
    $^1$ Steward Observatory, University of Arizona, 933 N. Cherry Avenue, 85721 Tucson, AZ, USA\\
	$^2$ DARK, Niels Bohr Institute, University of Copenhagen, Denmark.\\
	$^3$ Department of Physics and Astronomy, University of California, Los Angeles, CA, 90095, USA\\
	$^4$ Department of Physics and Astronomy, University of Southampton, Highfield, Southampton, SO17 1BJ, UK\\
	$^5$ Institute of Astronomy and Astrophysics, Academia Sinica,  Roosevelt Rd,  Taipei 10617,  Taiwan,  R.O.C.\\
	$^6$ Retired\\
	$^7$ Department of Physics, Geology, and Engineering Technology, Northern Kentucky University, 1 Nunn Dr., Highland Heights,\\ 
	KY 41099\\
	$^8$ Ohio State University, McPherson Laboratory, 140 West 18th Avenue, Columbus, Ohio, USA}
    
\begin{document}
	
\maketitle

\begin{abstract}
Mrk 590 is a known changing-look AGN which almost turned off in 2012, and then in 2017 partially re-ignited into a repeat flaring state, unusual for an AGN. Our \emph{Swift} observations since 2013 allow us to characterise the accretion-generated emission and its reprocessing in the central engine of a changing-look AGN. The X-ray and UV variability amplitudes are higher than those typically observed in `steady-state' AGN at similar moderate accretion rates; instead, the variability is similar to that of highly accreting AGN. The unusually strong X-ray to UV correlation suggests that the UV-emitting region is directly illuminated by X-ray outbursts. We find evidence that the X-rays are reprocessed by two UV components, with the dominant one at $\sim$3 days and a faint additional reprocessor at near-zero lag. However, we exclude a significant contribution from diffuse broad line region continuum, known to contribute for bona-fide AGN. A near-zero lag is expected for a standard `lamp-post' disk reprocessing model with a driving continuum source near the black hole. That the overall UV response is dominated by the $\sim$3-day lagged component suggests a complicated reprocessing geometry, with most of the UV continuum not produced in a compact disk, as also found in recent studies of NGC 5548 and NGC 4151. Nonetheless, the observed flares display characteristic timescales of $\sim$100  rest-frame days, consistent with the expected thermal timescale in an accretion disk.
\end{abstract}


\section{Introduction}\label{sec:introduction}

Active Galactic Nuclei (AGN) emit brightly across the entire electromagnetic spectrum, from radio waves to X-rays and Gamma radiation \citep[e.g.,][]{Elvis1994,Richards2006}. Their bright, blue UV--optical continua are believed to be emitted by a thermal accretion disk around a supermassive black hole \citep[e.g.,][]{Shakura1973,Novikov1973}, while a hot, optically thin corona emits the observed X-ray continuum \citep[e.g.,][]{Haardt1993,Petrucci2000,Lusso2016}. For the standard thermal accretion disk models, large variations in the accretion flow that ultimately powers the UV--optical emission should occur on the viscous timescale of the disk, which for AGN is $>10^{5}$ years \citep[e.g.,][]{Noda2018}. The observed strong UV variability, on timescales of days to months for lower-luminosity AGN \citep[e.g.,][]{Collier2001,Kelly2009,Cackett2015,McHardy2018}, is typically attributed to reprocessing of variable X-ray emission \citep[e.g.,][]{Collier2001,Kelly2009,Cackett2015,McHardy2018}. In this `lamp-post' scenario \citep[e.g.,][]{Collier1999,Cackett2007}, the hot and compact X-ray corona is situated near the central black hole and illuminates the cooler, optically thick accretion disk. Alternatively, disk instabilities not captured by the standard model may explain the rapid UV variability \citep[e.g.,][]{Collier2001, Hameury2009,Noda2018,Jiang2020}.

In recent years, several extreme AGN variability events have been observed \citep[e.g.,][]{Penston1984,Denney2014,Runnoe2016,LaMassa2017,MacLeod2019,Rumbaugh2018}. These so-called changing look AGN (CLAGN) are characterized by the appearance or disappearance of the UV--optical continuum and broad emission line spectral components, on timescales of months to years. This corresponds to a transition between AGN spectra that contain broad emission lines (i.e., Seyfert 1-type) and those with only narrow lines (Seyfert 2-type). For non-CLAGN, these distinct observational classes can be explained by bulk obscuration along some lines of sight \citep[e.g.,][]{Antonucci1993}. A few CLAGN are indeed consistent with variable absorption along our line of sight to the continuum source \citep[e.g.,][]{Goodrich1989,Goodrich1995}, while the remainder are likely due to changes in the luminosity of the ionizing continuum \citep[e.g.,][]{Penston1984,Denney2014,Runnoe2016,Noda2018,Kynoch2019}. CLAGN events due to strong continuum variability represent a challenge to the standard disk models, as they occur on timescales much shorter than the viscous timescale of the disk \citep{Noda2018,Lawrence2018,Dexter2019}. 
Suggested mechanisms to produce such strong and rapid variability include an unstable transition between a `puffed-up' inner advective disk and a geometrically thin, thermal outer disk \citep{Sniegowska2020,Pan2021}, and disk density inversions due to Hydrogen or Iron opacity fronts \citep{Jiang2020}.

Mrk 590 is a nearby AGN with a black hole mass $M_\mathrm{BH}=4.75(\pm0.74)\times10^7  M_\odot$, as determined via reverberation mapping \citep{Peterson2004}. During the 1980s and 1990s, this source displayed a typical Seyfert 1 UV-optical spectrum, including the AGN continuum component and broad H$\beta$, C \Rom{4} and Ly$\alpha$ emission. \citet{Denney2014} report a gradual decline in the continuum and broad emission line fluxes between 1989 and 2013. In particular, their 2014 optical (3500 \AA--7200 \AA) spectrum is consistent with host galaxy emission plus AGN narrow-line emission, displaying no evidence of AGN continuum or broad H$\beta$ emission. Similarly, the UV continuum at 1450 \AA\, decreased by a factor $\sim100$ between 1991 and 2013, while the broad components of the C \Rom{4} and Ly$\alpha$ lines disappear (or are severely diminished) over the same time period. Based on analysis of their 2013 \emph{Chandra} 0.5-10 keV X-ray observation, \citet{Denney2014} do not find evidence for an increase in intrinsic absorbing column density in the low-flux state. Instead, they suggest that the AGN `turned off' in terms of its UV--optical continuum emission. While the narrow emission lines do not disappear, \citet{Denney2014} report a fainter narrow emission line flux in 2013--2014 than in earlier observations. This is consistent with the `turn-off' scenario: the narrow-line emitting region is more extended than the broad-line region, and has longer recombination times due to its lower density. It will therefore respond only gradually to the diminishing flux of ionizing continuum photons. The radio variability displayed by Mrk 590 between 1995-2015 also supports an accretion rate change instead of line-of-sight obscuration driving the variability \citep{Koay2016}. ALMA observations reveal a circumnuclear ring of molecular gas, along with a kinematically disturbed clump of gas $\sim200$ pc from the nucleus that may intermittently feed the AGN on long timescales \citep{Koay2016b}. It is unclear whether these disturbed large-scale gas dynamics are connected to the short-term changing-look events. Mrk 590 displays a variable soft X-ray excess emission component, which faded to an undetectable level between 2007--2011 \citep{Rivers2012}. However, \citet{Mathur2018} detect soft excess in \emph{Chandra} observations during the 2014 low-luminosity state. At this time the soft excess component, while faint in terms of X-ray flux, is brighter than expected given the UV -- soft excess luminosity relationship presented by \citet{Mehdipour2011}. They interpret this as an early indication that Mrk 590 was re-igniting; it may thus be a precursor of the strong X-ray and UV flaring behavior documented in this work.

To study the post turn-off evolution of Mrk 590 and gain insight into the underlying physics of its changing-look behavior, we initiated intermittent X-ray and UV-optical monitoring observations with \emph{Swift} XRT and UVOT starting December 2013. During 2017, we observed a factor $\sim5$ increase in the X-ray flux, over a timescale of a few months, with corresponding increases in the UV fluxes. The broad Balmer emission lines also reappeared at this time \citep{Raimundo2019}. Since the 2017 flare-up, we have monitored Mrk 590 at least every $\sim14$ days, with periods of more intense monitoring during 2017 and 2018. In this work, we present the \emph{Swift} monitoring observations spanning 2013--2021 (\S \ref{sec:results_lc}), and study the response of the UV--optical continuum to the X-ray variability (\S \ref{sec:results_rm}). We discuss our results in \S \ref{sec:discussion}, and conclude in \S \ref{sec:conclusion}. In future work we will investigate the accretion physics in the high- and low-luminosity states through broad-band SED modeling, test whether the flares since 2017 exhibit any periodicity, and present an analysis of the evolution of the X-ray emission and reflection spectra.

\section{Observations}\label{sec:observations}

\begin{table*}
\caption{Individual \emph{Swift} XRT observations\label{tab:xrt_all}}
\begin{tabular}{ccccccccc}
\hline
\hline
MJD & Date & Observation & Exposure & 0.3-10 keV & $F_\mathrm{0.3-10}$ & $F_\mathrm{0.3-2}$ & $F_\mathrm{2-10}$ & $\Gamma_\mathrm{0.3-10}$\\ 
 & & ID & time [s] & counts &  &  &  & \\
(1) & (2) & (3) & (4) & (5) & (6) & (7) & (8) & (9) \\
\hline
54627 & 10/06/2008 & 00037590001 & 4465 & 514.4 & 5.75$_{-0.39}^{+0.30}$ & 2.21$_{-0.09}^{+0.12}$ & 3.53$_{-0.30}^{+0.30}$ & 1.64$_{-0.06}^{+0.06}$ \\
56636 & 12/10/2013 & 00037590002 & 1068 & 70.3 & 3.18$_{-0.42}^{+0.70}$ & 1.05$_{-0.14}^{+0.11}$ & 2.13$_{-0.34}^{+0.48}$ & 1.50$_{-0.17}^{+0.17}$ \\
56640 & 12/14/2013 & 00037590003 & 963 & 82.2 & 3.50$_{-0.44}^{+0.62}$ & 1.56$_{-0.15}^{+0.16}$ & 1.94$_{-0.37}^{+0.43}$ & 1.78$_{-0.16}^{+0.16}$ \\
\hline
\end{tabular}

{\raggedright The \emph{Swift} data analyzed in this work are observed during 10th December 2013 (Modified Julian Date, MJD, 56636) to 4th March 2021 (MJD 59277). For completeness, we also include the earliest \emph{Swift} observation of Mrk 590 (MJD 54627, 10th June 2008) here. All uncertainties represent 90\% confidence intervals. We present the first three table entries here for guidance; the full version is available in the online version of this article. \\
	Columns: \emph{(1)} Modified Julian Date (MJD), i.e., the number of days since November 17th, 1858.
	\emph{(2)} Calendar date (MM/DD/YYYY).
	\emph{(3)} \emph{Swift} Observation ID. 
	\emph{(4)} XRT on-source exposure time. 
	\emph{(5)} Background-subtracted XRT counts in the energy range 0.3-10 keV. 
	\emph{(6)} Integrated 0.3--10 keV flux, in units of 10$^{-12}$ erg cm$^{-2}$ s$^{-1}$. 
	\emph{(7)} Integrated 0.3--2 keV flux, in units of 10$^{-12}$ erg cm$^{-2}$ s$^{-1}$. 
	\emph{(8)} Integrated 2--10 keV flux, in units of 10$^{-12}$ erg cm$^{-2}$ s$^{-1}$. 
	\emph{(9)} Photon index for 0.3--10 keV model fit.\par}
\end{table*}

We observed Mrk 590 with the Neil Gehrels Gamma-Ray Burst Explorer mission \emph{Swift} \citep{Gehrels2004} intermittently since 2013. Following a sharp rise in the X-ray flux in August 2017, we obtained roughly bi-weekly observations (\emph{Swift} GO Cycle 14, Programs 1417159 and 
1417168 - PI: Vestergaard; joint \emph{NuSTAR}--\emph{Swift}, \emph{NuSTAR} Cycle 5, Program 5252 - PI: Vestergaard), with an additional period of high-cadence (1-2 days) monitoring during September 2017 -- February 2018 (\emph{Swift} Cycle 13 Director's Discretionary Time, PI: Vestergaard). Listed in order of the first observation, the \emph{Swift} target IDs for the data presented here are 37590, 80903, 88014, 94095, 10949, 11481, 11542, and 13172. In total, \emph{Swift} performed 198 individual observations of Mrk 590 up to 4th March 2021. \emph{Swift} observes simultaneously with the UltraViolet and Optical Telescope \citep[UVOT,][]{Roming2005} and the X-Ray Telescope \citep[XRT,][]{Burrows2005} instruments, with a single UVOT imaging filter in operation at any one time. The individual observation IDs and XRT exposure times are listed in Table \ref{tab:xrt_all}. Mrk 590 is behind the Sun, and thus unobservable with \emph{Swift}, from early March through early June. A supernova was detected in the host galaxy during July 2018; we discuss the influence of this event on our flux measurements in \S \ref{sec:results_lc}.
\paragraph*{\emph{Swift} UVOT:} Prior to February 2020, we observe with all UVOT imaging filters, using an exposure time distribution of 1:1:1:2:3:4 for \emph{V}, \emph{B}, \emph{U}, \emph{UVW1}, \emph{UVM2} and \emph{UVW2}, respectively. We choose this exposure time distribution as the AGN UV emission can be very faint for CLAGN in low-luminosity states, and is most easily detected in the far-UV. Since February 2020, we no longer observe with \emph{UVM2}, preferring instead to obtain deeper imaging in the other far-UV filters.
\paragraph*{\emph{Swift} XRT:} We observe with the XRT in photon counting (PC) mode \citep{Hill2004}. We verify that the observations are not affected by photon pile-up during any observations (\S \ref{sec:data_xrt}). Our typical XRT monitoring observations have exposure times of $\sim$2 ks, which for Mrk 590 is sufficient to determine the overall flux and X-ray continuum photon index. 

\section{Data Processing}\label{sec:data}

\subsection{\emph{Swift} UVOT photometric extraction}\label{sec:data_uvot}

We process the UVOT data using the standard pipeline tools provided as part of the  \emph{HEASoft} package\footnote{URL: \url{https://heasarc.gsfc.nasa.gov/lheasoft/}}. The UVOT detector suffers from small-scale sensitivity issues, as identified by \citet{Edelson2015} and subsequently documented in the CALDB release note SWIFT-UVOT-CALDB-17-01\footnote{URL: \url{SWIFT-UVOT-CALDB-17-01}}. The affected detector regions depend on the applied imaging filter. Using the provided small-scale sensitivity maps, we identify observations for which the source region is affected, and discard these observations from our analysis. Of the 198 observations in each imaging filter, we discard 6 observations using \emph{UVW1}, five \emph{UVM2} observations, and four \emph{UVW2} observations. The \emph{U}, \emph{B} and \emph{V} bands are not affected by the small-scale sensitivity issue.

We combine the individual UVOT snapshots for each observation using the standard pipeline processing for imaging mode (\textsc{HEASoft} version 6.26.1 or above, UVOTA CALDB version 20170922). We extract source and background fluxes from the resulting images using the \emph{`uvotsource'} task, setting \emph{`aprcorr=curveofgrowth'}. We use a circular source extraction aperture with a radius of 3 arcseconds, as recommended by the \emph{`uvotsource'} documentation\footnote{URL: \url{https://heasarc.gsfc.nasa.gov/ftools/caldb/help/uvotsource.html}}, and positioning the background region on blank sky in the same detector quadrant as the source. The \emph{`uvotsource'} task converts the observed count-rates to flux densities at the filter central wavelength, assuming a mean GRB spectrum \citep{Poole2008,Breeveld2010}. As we only study the variability behavior in the present work, we include only the statistical uncertainties in our error budget, ignoring the photometric calibration uncertainty. We correct the flux densities for Galactic reddening assuming $E(B-V)=0.0246\pm0.0005$ \citep{Schlafly2011} and using an O'Donnell reddening law with $R_\nu=3.1$ \citep{O'Donnell1994}. We list the flux densities for each observation in Table \ref{tab:uvot_all}. 

\begin{table*}
\caption{Individual \emph{Swift} UVOT observations.\label{tab:uvot_all}}
\begin{tabular}{cccccccc}
\hline
\hline
MJD & Date & $F_\mathrm{V}$ & $F_\mathrm{B}$ & $F_\mathrm{U}$ & $F_\mathrm{UVW1}$ & $F_\mathrm{UVM2}$ & $F_\mathrm{UVW2}$ \\
(1) & (2) & (3) & (4) & (5) & (6) & (7) & (8)\\
\hline
54627 & 10/06/2008 & 4.32$\pm0.15$ & 2.92$\pm0.10$ & 1.23$\pm0.05$ & 0.88$\pm0.05$ & 0.74$\pm0.04$ & 4.32$\pm0.05$ \\
56636 & 12/10/2013 & 4.16$\pm0.19$ & 2.96$\pm0.13$ & 1.21$\pm0.07$ & 0.60$\pm0.05$ & 0.57$\pm0.05$ & 4.16$\pm0.05$ \\
56640 & 12/14/2013 & 4.31$\pm0.20$ & 2.89$\pm0.13$ & 1.24$\pm0.07$ & 0.70$\pm0.06$ & 0.56$\pm0.06$ & 4.31$\pm0.06$ \\
\hline
\end{tabular}

{\raggedright The \emph{Swift} data analyzed in this work are observed during 10th December 2013 (MJD 56636) to 4th March 2021 (MJD 59277). For completeness, we also include the first \emph{Swift} observation of Mrk 590 (MJD 54627, 10th June 2008) here. We present the first three table entries here for guidance; the full version is available in the online version of this article.\\
Columns: \emph{(1)} Modified Julian Date. \emph{(2)} Calendar date (MM/DD/YYYY).
\emph{(3)} to \emph{(7)}: \emph{Swift} UVOT flux density in the filters $V$ to $UVW2$, assuming a power-law SED within the filter bandpass. These flux densities are corrected for Galactic dust reddening (\S \ref{sec:data_uvot}). Units of 10$^{-15}$ erg cm$^{-2}$ \AA$^{-1}$ s$^{-1}$. \par}

\end{table*}

\subsection{\emph{Swift} XRT data processing and spectral modeling}\label{sec:data_xrt}

We process the XRT PC mode event files using the standard pipeline software (\textsc{HEASoft} version 6.26.1 or above), using the \emph{`xselect'} task to prepare source and background `.pha' files for analysis. We set the \emph{`xselect'} grading threshold to 0--12, also discarding events with energies outside the 0.3-10 keV XRT sensitivity range, and events taking place outside the `good time intervals' defined by the observation logs (e.g., taking place while the spacecraft is slewing). We use a circular source extraction region with a radius of 20 pixels (47 arcseconds), which encloses $\sim90$\% of the on-axis point spread function at 1.5 keV \citep{Moretti2004}. We extract a circular background region of radius 216 arcseconds, positioned to avoid an additional faint X-ray source at RA: 2:14:35.3, Dec: -0:42:44.6. We generate Auxiliary Response Files (ARFs) for each observation using the task \emph{`xrtmkarf'}. These files include information on the effective area, quantum efficiency, and PSF profile for a given observation, and are used in the spectral analysis. While the majority of our observations consist of a single telescope pointing, we use XSELECT to combine observations in cases where the exposure time is split over two or more pointings.

For observations with 0.3-10 keV count-rates exceeding $0.5$ cts s$^{-1}$, we test for the effects of photon pile-up by modelling the observed azimuthally averaged point spread function as a King profile, excluding the inner 10 arcseconds. In all cases, an extrapolation of this model to the central region confirms that the point spread function core is consistent with the King profile. Thus, our XRT observations are not affected by pile-up.

We model each individual XRT spectrum as a power-law continuum plus Galactic absorption, using the Galactic absorption column density towards Mrk 590, $N_\mathrm{H,Gal}=2.77\times10^{20}$ cm$^{-2}$ \citep{Bekhti2016}. The free parameters of this model are the photon index $\Gamma_\mathrm{XRT}$, and the flux normalization at 1 keV. We find that the simple power-law models are fully consistent with the data for the individual $\sim2$ ks observations presented here. While a soft X-ray excess is present in Mrk 590 in 2004 \citep{Longinotti2007} and in the 2014 low-luminosity state \citep{Mathur2018}, our individual \emph{Swift} monitoring observations do not require a soft excess component. We will present a detailed analysis of the X-ray emission spectrum of Mrk 590, based on stacked \emph{Swift} XRT data and on recent and on-going \emph{XMM-Newton} and \emph{NuSTAR} observations, in future work. We extract the full-band (observed-frame 0.3-10 keV), soft-band (0.3-2 keV) and hard-band (2-10 keV) X-ray fluxes from these models. All X-ray fluxes presented in this work are corrected for Galactic absorption.

\section{\emph{Swift} XRT and UVOT lightcurves}\label{sec:results_lc}

\begin{figure*}
    \newcommand{\picscale}{0.35}
    \centering
    \includegraphics[scale=\picscale,trim={0 20 0 0}, clip]{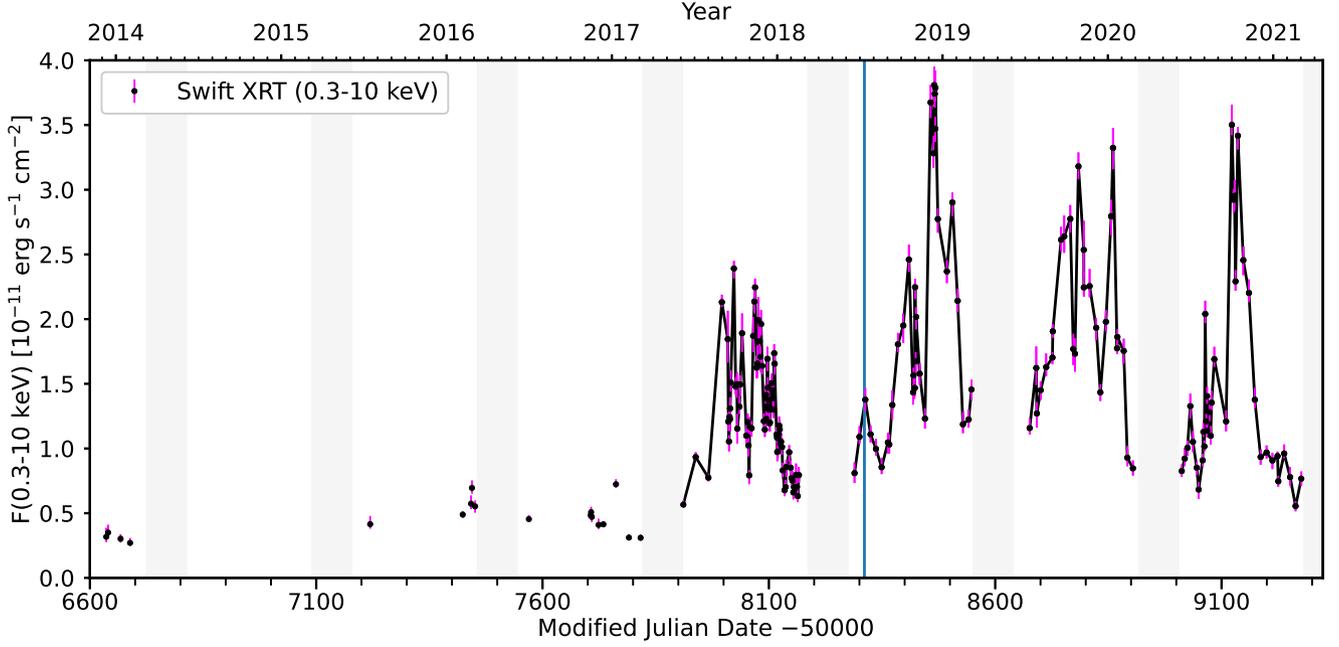}
    \caption{\emph{Swift} XRT lightcurve for the period December 2013 -- March 2021. The XRT lightcurve (\emph{black dots}) provides the absorption-corrected integrated flux between 0.3 - 10.0 keV, based on a power-law continuum model and corrected for Galactic absorption. Magenta error bars represent the 90\% confidence interval on the model flux. For the higher-cadence observations since 2017, we connect data points with a black line to guide the eye. The vertical blue line indicates the detection date of the supernova ASASSN-18pb (\S \ref{sec:results_lc}). Mrk 590 is unobservable with \emph{Swift} between $\sim$5th March -- early June each year, as it is behind the Sun (light-gray shaded regions).}
    \label{fig:xrt_lightcurve}
\end{figure*}

\begin{figure}
    \centering
    \includegraphics[scale=0.21]{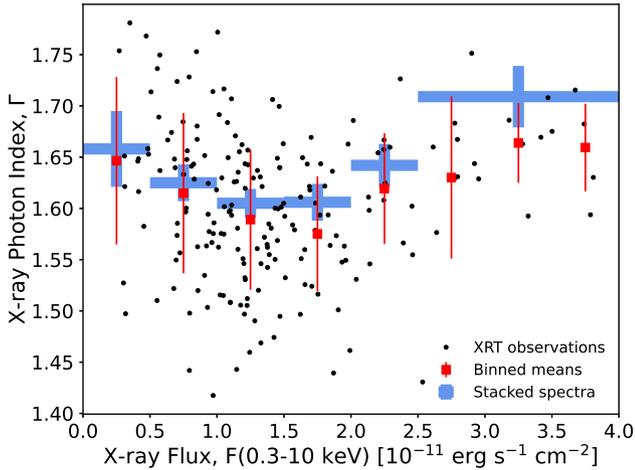}
    \caption{Our individual \emph{Swift} XRT observations display a large scatter in the X-ray photon index $\Gamma$. The uncertainties on $\Gamma$ are of order $\pm0.07$; for clarity, we do not show errors on the individual measurements (\emph{black points}). We uniformly bin the individual measurements by 0.3--10 keV flux (bin-width $5\times10^{-12}$ erg cm$^{-2}$ s$^{-1}$) and show the mean $\Gamma$ values (\emph{red squares}) and their standard deviations \emph{(red error-bars)}. We also stack the XRT spectra binned by 0.3-10 keV flux (\emph{blue crosses; horizontal `error bars' illustrate the flux bins used for stacking}). The $\Gamma$ values measured for the stacked spectra are in agreement with the binned mean values to within the sample standard deviations. However, the stacking analysis indicates a softening of the X-ray spectra when the flux exceeds $F_{\mathrm{0.3-10}}\gtrsim2.5\times10^{-11}$ erg cm$^{-2}$ s$^{-1}$, and perhaps also at the lowest X-ray fluxes, suggesting a `U-shaped' trend with $F_{\mathrm{0.3-10}}$.}
    \label{fig:xrt_Gamma}
\end{figure}

\paragraph*{X-ray lightcurves:} We present the integrated fluxes (full-band $F_{0.3-10}$, soft-band $F_{0.3-2}$ and hard-band $F_{2-10}$), and the 0.3-10 keV photon index $\Gamma_{0.3-10}$, for our power law model fits to the individual \emph{Swift} XRT observations in Table \ref{tab:xrt_all}. The X-ray flux appears to remain at a low level between 2013 -- early 2017, although our time sampling is sparse during this period. In late August 2017 we observe an abrupt flare-up, with $F_{0.3-10}$ increasing by a factor $\sim5$ relative to the January 2017 level. (Figure \ref{fig:xrt_lightcurve}). After this initial flare-up, the X-ray emission is variable on timescales of days to weeks, with prominent flare-ups and subsequent fading occurring during each observing season since 2017. The maximum observed X-ray flux, $F_{\mathrm{0.3-10}}=3.8\times10^{-11}$ erg cm$^{-2}$ s$^{-1}$, occurred in December 2018. In February 2021, the X-ray flux reached its lowest level since 2017, $F_{\mathrm{0.3-10}}=5.6\times10^{-12}$ erg cm$^{-2}$ s$^{-1}$. 

\paragraph*{X-ray spectral variability:} The measured X-ray photon indices $\Gamma$ for our individual observations display a large scatter at a given X-ray flux level, with $1.4\lesssim\Gamma\lesssim1.8$ (Figure \ref{fig:xrt_Gamma}). The modeling uncertainties on $\Gamma$ are of order $\pm0.1$ for these short observations (Table \ref{tab:xrt_all}). To test for an underlying dependence of the spectral shape on $F_{\mathrm{0.3-10}}$, we firstly determine the average $\Gamma$ values for the individual observations, in flux bins of width 5$\times10^{-12}$ erg cm$^{-2}$ s$^{-1}$ (Figure \ref{fig:xrt_Gamma}, red squares). Secondly, we stack all observations within a flux bin, and model the stacked spectra as a power-law plus Galactic absorption. We use the Swift XRT Data Products Generator \citep{Evans2009} to stack the spectra, and fit them with the same model as used for our individual observations. The smaller uncertainties on $\Gamma$ for the stacked spectra (Figure \ref{fig:xrt_Gamma}, blue crosses) reveal a significant softening of the X-ray spectrum at the highest observed flux levels, $F_{\mathrm{0.3-10}}\gtrsim2.5\times10^{-11}$ erg cm$^{-2}$ s$^{-1}$. We also see hints of spectral softening at low flux levels, although the change in $\Gamma$ is modest relative to the measurement uncertainties. If real, this `U-shaped' trend provides further evidence that Mrk 590 is fluctuating near an important luminosity threshold during these observations. Low-luminosity AGN typically display `harder-when-brighter' behavior, \emph{i.e.,} smaller $\Gamma$ at higher luminosity, while brighter Seyferts and quasars display a `softer-when-brighter' trend. The transition between the two trends is typically identified using statistical samples of AGN \citep[e.g.,][]{Gu2009,Connolly2016}. However, recent studies find that CLAGN tend to accrete near this threshold \citep[e.g.,][]{Ai2020}, and in some cases individual sources transition from `harder-when-brighter' behavior to the opposite \citep[e.g.,][]{Xie2016,Lyu2021,Liu2022}. We will address this issue further in future work, based on deeper X-ray observations at both low and high luminosities.

\begin{figure*}
     \newcommand{\picscale}{0.42}
     \centering
     \includegraphics[scale=\picscale,trim={90 70 70 80}, clip]{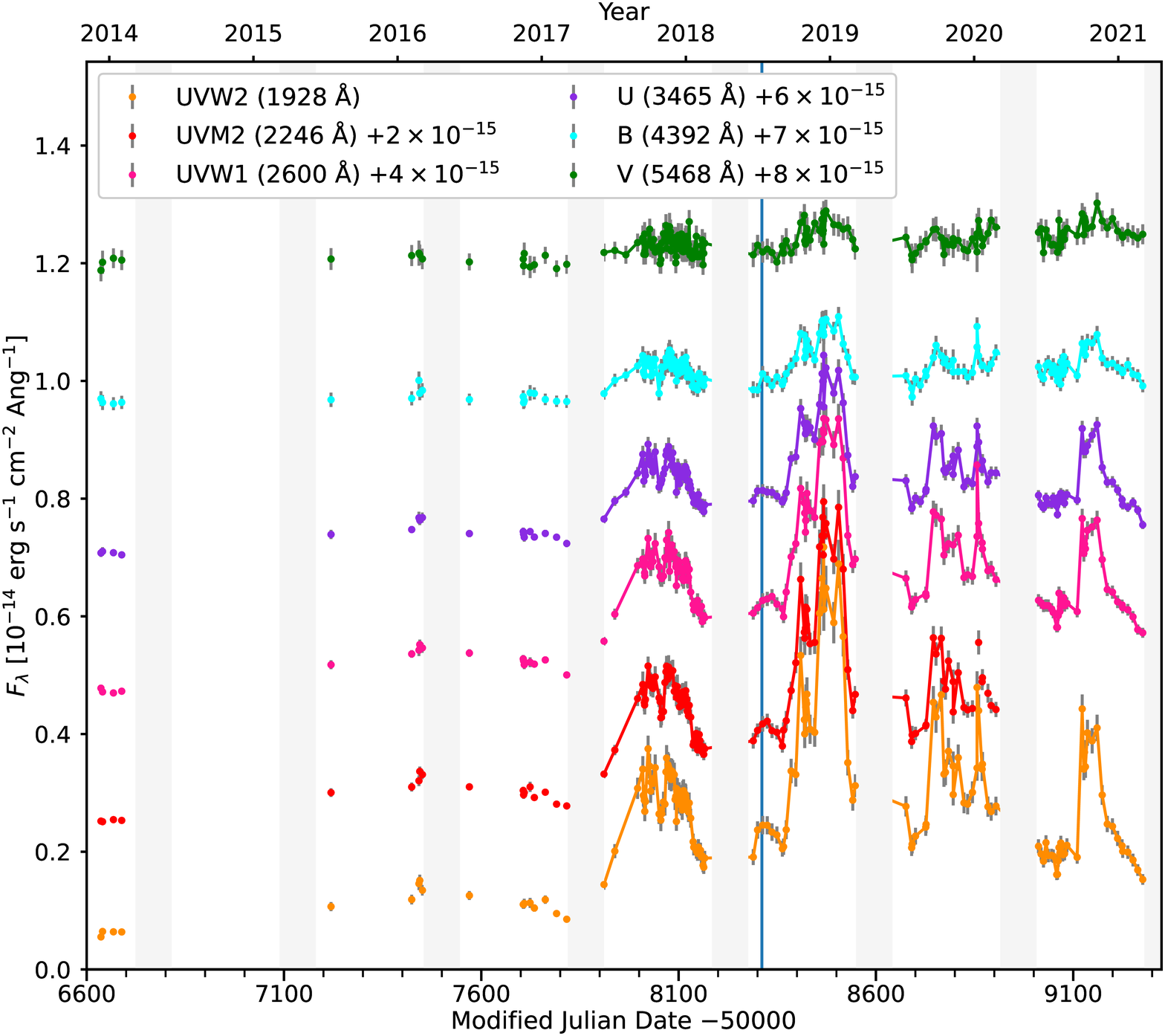}
     \caption{\emph{Swift} UVOT lightcurves for the period 10th December 2013 -- 4th March 2021. The flux densities are corrected for Galactic reddening, but are not corrected for host galaxy emission. The lightcurves are shifted by an arbitrary constant flux for presentation purposes. Each UVOT lightcurve provides the flux density at the central wavelength of each filter, assuming a power-law SED. The gray errorbars represent $1\sigma$ photometric uncertainties derived from the \emph{`uvotextract'} task. For the higher-cadence observations since 2017, we connect data points with colored lines to guide the eye. Mrk 590 is behind the Sun and therefore unobservable with \emph{Swift} between early March -- early June each year (light-gray shaded regions). The vertical blue line indicates the detection date of the supernova ASASSN-18pb (\S \ref{sec:results_lc}).}
     \label{fig:lightcurves}
\end{figure*}

\paragraph*{UV--optical lightcurves:} The far--UV and $U$ bands display flares roughly concurrently with those observed in X-rays (Figure \ref{fig:lightcurves}; Table \ref{tab:uvot_all}). The brightest UV--optical flare occurred during December 2018. The $UVW2$ flux varies by a factor $\sim$12.8 between the 2014 low-flux state and its 2018 peak flux. In \S \ref{sec:discussion_flares} we argue that the observed UV flares are likely due to the re-ignition of the AGN continuum component.

\paragraph*{A supernova eruption during July 2018:} The supernova ASASSN-18pb erupted in the host galaxy of Mrk 590, with a peak SDSS $g$ magnitude $\sim16.8$, around 12th July 2018 \citep{Atel2018}. We illustrate the detection date with vertical blue lines in Figures \ref{fig:xrt_lightcurve} and \ref{fig:lightcurves}. Here, we argue that the minor X-ray and UV flares that occur roughly concurrently are \emph{unrelated} to the supernova event. On 14th July 2018 we detect the supernova as a point source at an angular separation of 7.8'' from the nucleus, with a flux of $4\times10^{-16}$ erg cm$^{-2}$ s$^{-1}$ \AA$^{-1}$ in $UVW2$. The resulting scattered light in our source extraction aperture is of order $10^{-17}$ erg cm$^{-2}$ s$^{-1}$ \AA$^{-1}$ in $UVW2$, according to the instrumental Curve of Growth for UVOT\footnote{SWIFT-UVOT-CALDB-104, \url{https://heasarc.gsfc.nasa.gov/docs/heasarc/caldb/swift/docs/uvot/uvot_caldb_psf_02.pdf}}. This scattered light is too faint to produce the observed UV flare. In the X-rays, our source extraction region (radius 47'') does include the supernova position as derived from UVOT. However, it is unlikely that the concurrent X-ray flare-up is due to supernova emission, for the following reasons. \emph{1)} The X-ray flare coincides with the $UVW2$ flare, for which we exclude a significant supernova contribution. \emph{2)} We fit a King profile centered on the AGN in the July 14th X-ray image, and find it consistent with a point source. \emph{3)} ASASSN-18pb is identified as a Type \textrm{Ia} supernova \citep{Khlat2018}, which typically achieve X-ray luminosities of $L_X\sim10^{40}$ erg s$^{-1}$, a factor $\sim10^3$ fainter than that of Mrk 590.

\paragraph*{Excess variance:} To quantify the amount of variability displayed by Mrk 590, and facilitate comparisons to other AGN, we need to account for fluctuations due to measurement uncertainties. Following \citet{Rodriguez1997}, we define the fractional excess variance, $F_\mathrm{var}$, as

\begin{equation*}
    F_\mathrm{var}=\frac{\sqrt{\sigma_\mathrm{lc}^2-\delta^2}}{\langle f\rangle}.
\end{equation*}

\noindent Here, $\sigma_\mathrm{lc}^2$ denotes the flux variance of a given lightcurve, $\delta$ is the mean flux uncertainty, and $\langle f\rangle$ is the mean flux density. The uncertainty on $F_\mathrm{var}$ is given by

\begin{equation*}
    \sigma_{F_\mathrm{var}}=\frac{1}{F_\mathrm{var}}\sqrt{\frac{1}{2N}}\frac{\sigma_\mathrm{lc}^2}{\langle f\rangle^2},
\end{equation*}

\begin{figure}
    \centering
    \includegraphics[scale=0.53]{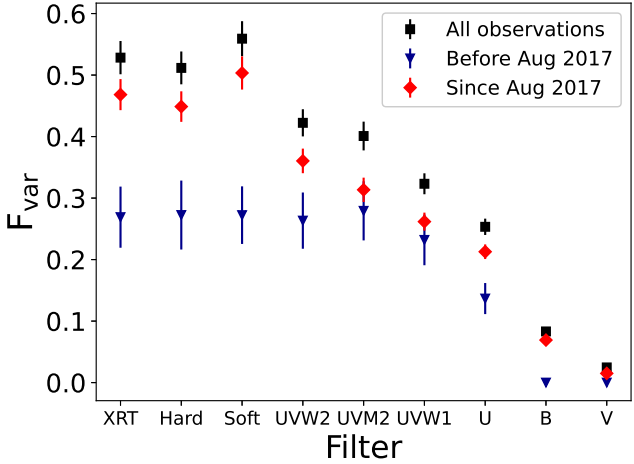}
    \caption{The normalized excess variability $F_\mathrm{var}$ calculated for the full-band (0.3--10 keV) X-ray lightcurve, the hard band (2--10 keV), the soft band (0.3--2 keV), and for each UVOT filter. We compare $F_\mathrm{var}$ for three time intervals. The full \emph{Swift} lightcurves observed since 2013 (\emph{black squares}) display the largest $F_\mathrm{var}$ in all bands, as this interval includes both the low-luminosity and the flaring state. The `flaring' epochs since August 2017 (\emph{red diamonds}) generally display higher excess variance than the observations prior to August 2017 (\emph{blue triangles}). We note that the sparse time sampling prior to August 2017 will suppress the measured $F_\mathrm{var}$ for that interval to some degree.}
    \label{fig:fvar}
\end{figure}

\noindent\citep{Edelson2002}, where $N$ is the number of observations in the lightcurve. For the full 0.3--10 keV X-ray lightcurves we find $F_\mathrm{var}=0.53\pm0.03$. The $F_\mathrm{var}$ values for the 0.3--2 keV and 2--10 keV lightcurves are consistent within the 1$\sigma$ uncertainties (Figure \ref{fig:fvar}; Table \ref{tab:fvar}). We also provide separate $F_\mathrm{var}$ values before and after the onset of flaring activity. For the X-ray lightcurves we measure $F_\mathrm{var}=0.47\pm0.03$ since August 1st 2017, compared to $F_\mathrm{var}=0.27\pm0.05$ prior to that date. While the sparse time sampling prior to the flare-up may suppress the measured $F_\mathrm{var}$ before 2017, our data are consistent with increased variability during the flares.

The UV lightcurves display $F_\mathrm{var}\sim30$\% after the initial flare-up (Table \ref{tab:fvar}; Figure \ref{fig:fvar}). We find a significant increase in $F_\mathrm{var}$ since August 2017 in the \emph{UVW2}, \emph{U} and \emph{B} filters. In general, $F_\mathrm{var}$ is larger at shorter wavelengths. This trend is also noted for other AGN \citep[e.g.,][]{Crenshaw1996,Fausnaugh2016,Gallo2018,Lobban2020}. It is expected both due to the well-known `bluer when brighter' behavior of AGN variability \citep[e.g.,][ and references therein]{Sun2014}, and due to dilution of the variable AGN emission by the host galaxy. Mrk 590 is an $Sa$-type spiral galaxy, with a dominant bulge component within the central 3'' sampled by our UVOT photometric aperture \citep{Bentz2006}. We expect a substantial constant emission component in the $B$ and $V$ filters due to this quiescent stellar population. The $V$ band displays near-zero excess variance, i.e., its observed variability is largely due to photometric uncertainty, although we see hints of a response to the 2018 flare (around MJD 58500, Figure \ref{fig:lightcurves}). Weak \emph{V} band responses are observed in several \emph{Swift} continuum reverberation mapping campaigns. They are partly due to a low signal-to-noise ratio in this filter, which can be mitigated in future monitoring campaigns by increasing the \emph{V} exposure times (Rick Edelson, \emph{priv.comm.}, 2022). We compare the excess variances of Mrk 590 during its flaring state with those observed for other AGN in \S \ref{sec:discussion_flares}.

\begin{table}
\caption{Fractional excess variances.\label{tab:fvar}}
\begin{tabular}{cccc}
\hline
\hline
	Bandpass & $F_\mathrm{var}$  & $F_\mathrm{var}$ & $F_\mathrm{var}$ \\
	 & 2013--2021 & Before Aug 2017 & Since Aug 2017 \\
	(1) & (2) & (3) & (4) \\
	\hline
	0.3--10 keV & 0.53$\pm0.03$ & 0.27$\pm0.05$ & 0.47$\pm0.03$ \\
	0.3--2 keV & 0.56$\pm0.03$ & 0.27$\pm0.06$ & 0.50$\pm0.03$ \\
	2--10 keV & 0.51$\pm0.03$ & 0.27$\pm0.05$ & 0.45$\pm0.02$ \\
	\emph{UVW2} & 0.42$\pm0.02$ & 0.26$\pm0.05$ & 0.36$\pm0.02$ \\
	\emph{UVM2} & 0.40$\pm0.02$ & 0.28$\pm0.05$ & 0.31$\pm0.02$ \\
	\emph{UVW1} & 0.32$\pm0.02$ & 0.23$\pm0.04$ & 0.26$\pm0.01$ \\
	\emph{U} & 0.25$\pm0.01$ & 0.14$\pm0.02$ & 0.21$\pm0.01$ \\
	\emph{B} & 0.08$\pm0.01$ & 0 & 0.07$\pm$0.01 \\
	\emph{V} & 0.02$\pm0.01$ & 0 & 0.01$\pm$0.01 \\
	\hline
	\end{tabular}
	\\

	Columns: (1) \emph{Swift} XRT energy range, or UVOT filter name.\\
	(2) Fractional excess variance $F_\mathrm{var}$ for the full \emph{Swift} lightcurve. \\
	(3) $F_\mathrm{var}$ for observations before August 2017, i.e., prior to the initial flare-up.\\
	(4) $F_\mathrm{var}$ for the observations since August 2017.\\
\end{table}

\paragraph*{Characteristic variability timescales:} To further quantify the variability behavior of Mrk 590 during the flares since 2017, we estimate the characteristic variability timescales of the X-ray and UV-optical lightcurves. On timescales of weeks to years, AGN UV--optical lightcurves can be described as stochastic processes governed by a characteristic timescale $\tau_\mathrm{char}$, beyond which the variability becomes uncorrelated \citep[e.g.,][]{Kelly2009,Kozlowski2010,MacLeod2010,Zu2013}. Here, we present estimates of  $\tau_\mathrm{char}$ for Mrk 590 based on two commonly applied methods: firstly using structure function analysis, and secondly, modeling the lightcurves as stochastic processes.

The structure function measures the variability power in a lightcurve as a function of a time delay $\tau$, and is in essence a time-domain equivalent of the power spectrum \citep[e.g.,][]{Hughes1992}. For stochastic processes governed by a single characteristic timescale, the amplitude of the structure function increases towards and flattens at the corresponding $\tau$ \citep[e.g.,][]{Collier2001ApJ}. We present the binned structure functions themselves, and describe our analysis approach, in Appendix \ref{sec:appendix_SF}. Our main findings are as follows. \emph{1)} The structure functions are roughly consistent with power laws for short time delays, and only display significant `flattening' upon reaching the long-term variance of the lightcurve. This is consistent with variability governed by a stochastic process. \emph{2)} The flattening of the structure functions suggests $\tau_\mathrm{char}\sim90$ rest-frame days for both the X-ray and UV--optical lightcurves, with no obvious dependence on wavelength (Table \ref{tab:tchar}). \emph{3)} The X-ray lightcurves display more variability power at short timescales than do the UV--optical lightcurves.

\begin{table}
\caption{Characteristic variability timescales, August 2017 -- March 2021.\label{tab:tchar}}
\begin{tabular}{ccc}
\hline
\hline
	Bandpass & $\tau_\mathrm{char}$  & $\tau_\mathrm{char}$  \\
	 & (Structure function) & (\textsc{Javelin})  \\
	(1) & (2) & (3) \\
	\hline

	0.3--10 keV & 90$^{+49}_{-28}$ & 24$_{-6}^{+10}$ (smoothed, 78$_{-25}^{+53}$) \\
	\emph{UVW2} & 85$^{+67}_{-35}$ & 111$_{-38}^{+92}$ \\
	\emph{UVM2} & 80$^{+84}_{-39}$ & 116$_{-41}^{+97}$ \\
	\emph{UVW1} & 83$^{+55}_{-31}$ & $122_{-43}^{+100}$ \\
	\emph{U} & 96$^{+61}_{-33}$ & 94$_{-32}^{+73}$ \\
	\emph{B} & 117$^{+307}_{-71}$ & 72$_{-25}^{+57}$ \\
	\hline
	\end{tabular}
	\\
    The \emph{V} band displays a near-zero excess variance, and is excluded from these analyses.\\
	Columns: (1) \emph{Swift} XRT energy range, or UVOT filter name.\\
	(2) Characteristic timescale derived from a power-law model fit to the structure function, as detailed in Appendix \ref{sec:appendix_SF}. Units of rest-frame days.\\
	(3) Characteristic timescale derived from \textsc{Javelin} modeling of each lightcurve. Units of rest-frame days. For the X-ray observations, we repeat the \textsc{Javelin} modeling after smoothing the lightcurve with a box-car width of 10 days to remove the high-frequency variability not typically seen in the UV--optical lightcurves. \\
\end{table}

To model the observed variability directly, we turn to the \textsc{Javelin} software package \citep{Zu2011}. \textsc{Javelin} models lightcurve variability as a first-order autoregressive process. These processes are suitable models of AGN variability when sampled at moderate cadences \citep[$\sim2$--10 days,][]{Kelly2009,Kozlowski2010,MacLeod2010,Zu2013}. A first-order autoregressive process with a characteristic timescale $\tau_\mathrm{char}$, irregularly sampled at times $t_1...t_N$, is given by

\begin{equation}\label{eq:ar1}
    f(t_i)=\exp\left(\frac{-(t_i-t_{i-1})}{\tau_\mathrm{char}}\right)f(t_{i-1})+\epsilon(t_i).
\end{equation}

Here, $f(t_i)$ is the flux measurement at time $t_i$. The quantity $\epsilon(t_i)$ is a random variable drawn from a Gaussian distribution with zero mean, and standard deviation $\sigma_\epsilon$ which governs the variability amplitude. \textsc{Javelin} then uses a Monte Carlo Markov Chain approach to determine the most likely values of $\tau_\mathrm{char}$ and of $\sigma_\epsilon$. For the UV--optical lightcurves, the $\tau_\mathrm{char}$ from \textsc{Javelin} modeling are consistent with those derived from our structure function analysis (Table \ref{tab:tchar}). In all cases we find $\tau_\mathrm{char}\sim100$ rest-frame days. We discuss some possible interpretations of these results in \S \ref{sec:discussion_flares}.

However, for the 0.3-10 keV X-ray lightcurve, \textsc{Javelin} finds a significantly shorter characteristic timescale, $\tau_\mathrm{char}=24_{-6}^{+24}$ rest-frame days. This is somewhat surprising. We certainly expect that the variability behavior \emph{on timescales of weeks to months} is similar for the X-ray and UV--optical, given that they display near-identical flaring patterns. As indicated by our structure function analysis (Appendix \ref{sec:appendix_SF}), the X-rays are more variable than the UV lightcurves on short timescales. Intriguingly, for a sample of AGN at $z<0.05$, \citet{Kelly2011} find that the optical lightcurves display a single characteristic timescale, whereas the X-ray lightcurves are better described by mixed processes governed by multiple characteristic timescales. We therefore speculate that the \textsc{Javelin} modeling is sensitive to an additional `rapidly flickering' component in the X-rays. To test this, we apply box-car smoothing to the X-ray lightcurve, with a smoothing width of 10 observed-frame days, in order to suppress the most rapid variability. We then repeat our \textsc{Javelin} modeling for this smoothed lightcurve, and find $\tau_\mathrm{char}=78_{-25}^{+53}$ rest-frame days. This measurement is fully consistent with both the X-ray structure function results, and with the UV--optical characteristic timescales.

\section{Continuum reverberation mapping analysis}\label{sec:results_rm}

\begin{figure*}
     \centering
     \includegraphics[scale=0.41,trim=70 70 70 70, clip]{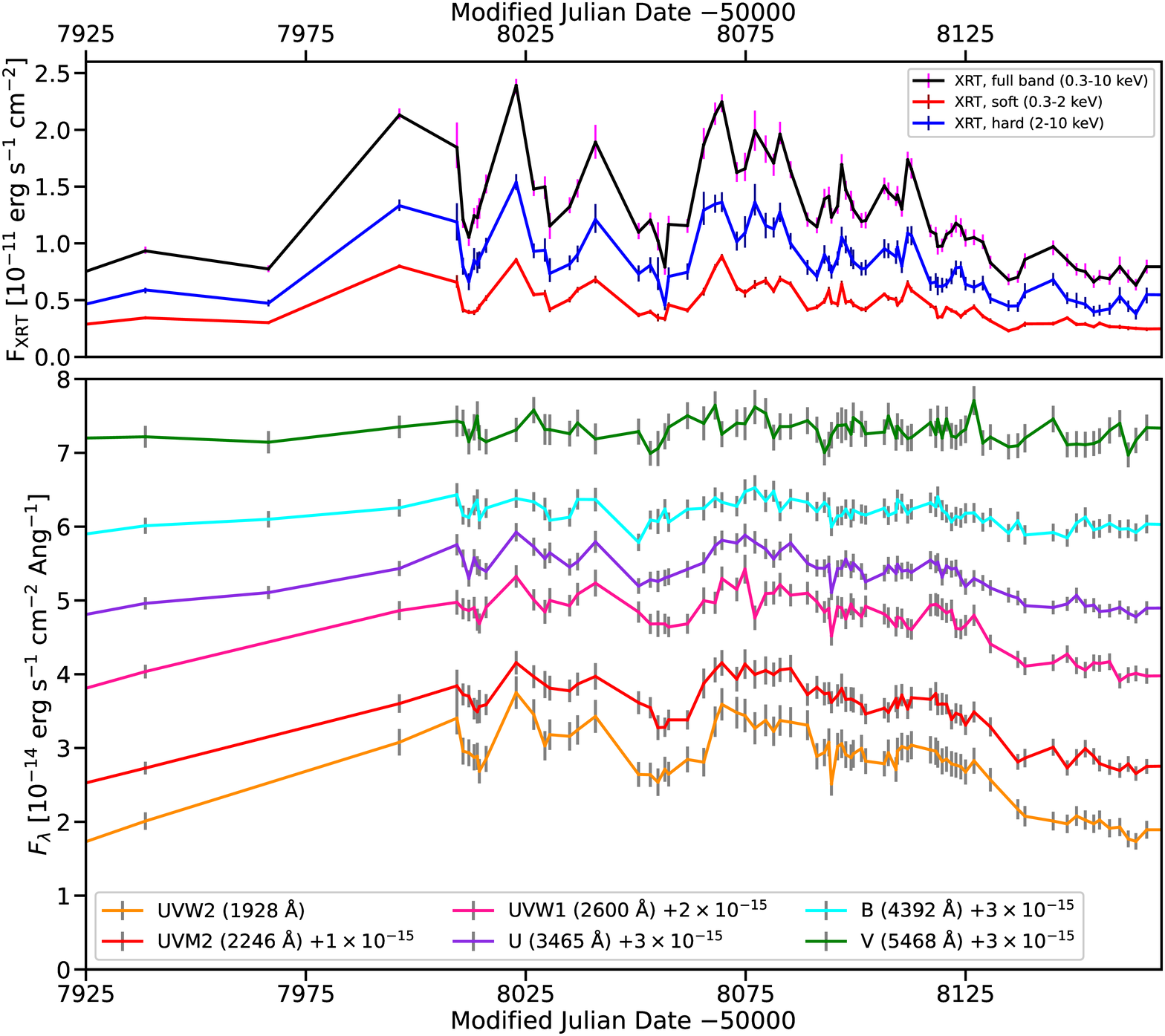}
     \caption{\emph{Swift} XRT (top) and UVOT (bottom) lightcurves for the period August 2017 -- February 2018. We use this interval for the reverberation mapping analysis (\S \ref{sec:results_rm}). The UVOT flux densities are corrected for Galactic reddening, but are not corrected for host galaxy emission. The UVOT lightcurves are shifted in flux by an arbitrary constant for presentation purposes.}
     \label{fig:lightcurves_2017}
\end{figure*}

For the period August 2017 -- February 2018 (i.e., immediately after the discovery of the first major flare-up) we obtained intensive \emph{Swift} monitoring, with an average observational cadence of $\sim2$ days during this period, including several short periods of roughly daily observations. We present `zoomed-in' lightcurves for this period in Figure \ref{fig:lightcurves_2017}. These data are suitable for continuum reverberation mapping, i.e., studying the response of the UV and optical continuum emission to variability in the inner regions of the accretion flow. Continuum reverberation mapping studies are currently available for a handful of low-redshift AGN \citep[e.g.,][]{Shappee2014,McHardy2014,Starkey2017,Edelson2019,Kara2021}. In order to robustly quantify the time delays and correlation strengths for our August 2017 -- February 2018 monitoring data, we apply two different reverberation mapping analysis methods: the interpolated cross correlation functions (\S \ref{sec:iccf}), and the \textsc{Javelin} method (\S \ref{sec:javelin}). We also investigate the dependence of the reverberation signal on source variability frequency, decomposing our lightcurves into slowly- and rapidly-varying components (\S \ref{sec:results_rm_filtering}).

\subsection{Interpolated cross-correlation functions}\label{sec:iccf}

\begin{figure*}
   \newcommand{\picscale}{0.52}
   \centering
   \includegraphics[scale=\picscale,trim=0 30 0 0, clip]{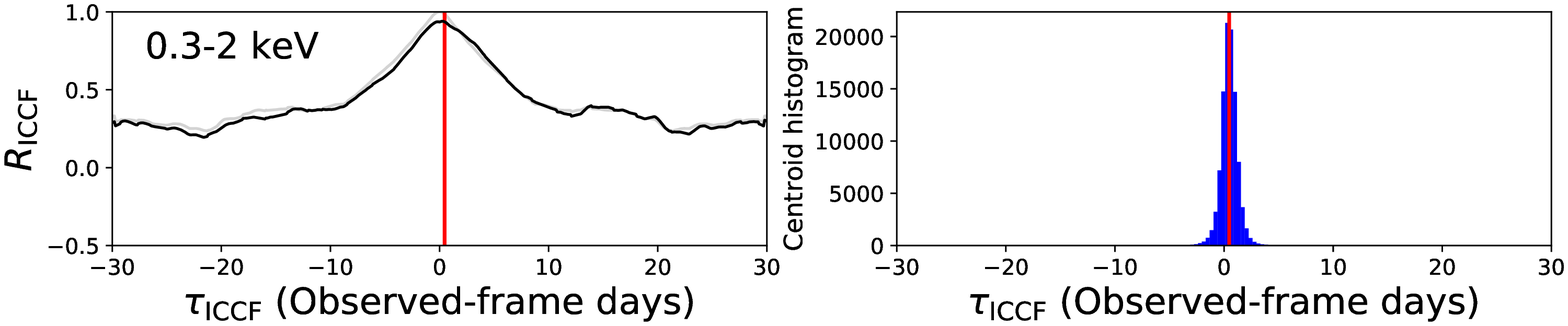}
   \includegraphics[scale=\picscale,trim=0 30 0 0, clip]{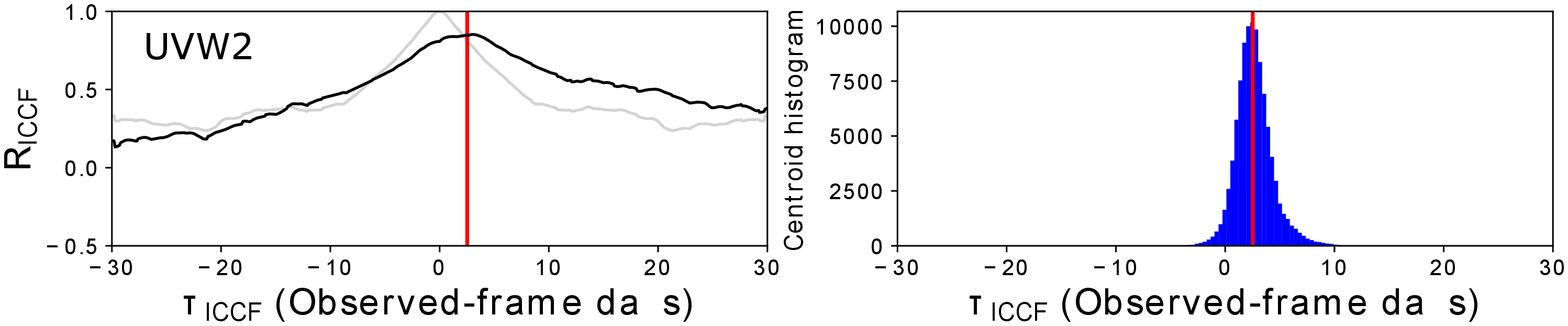}
   \includegraphics[scale=\picscale,trim=0 30 0 0, clip]{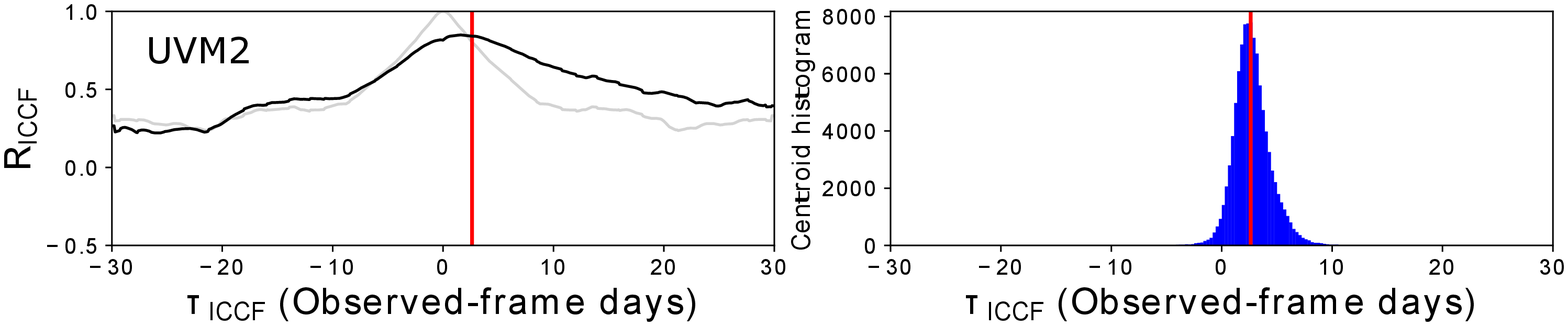}
   \includegraphics[scale=\picscale,trim=0 30 0 0, clip]{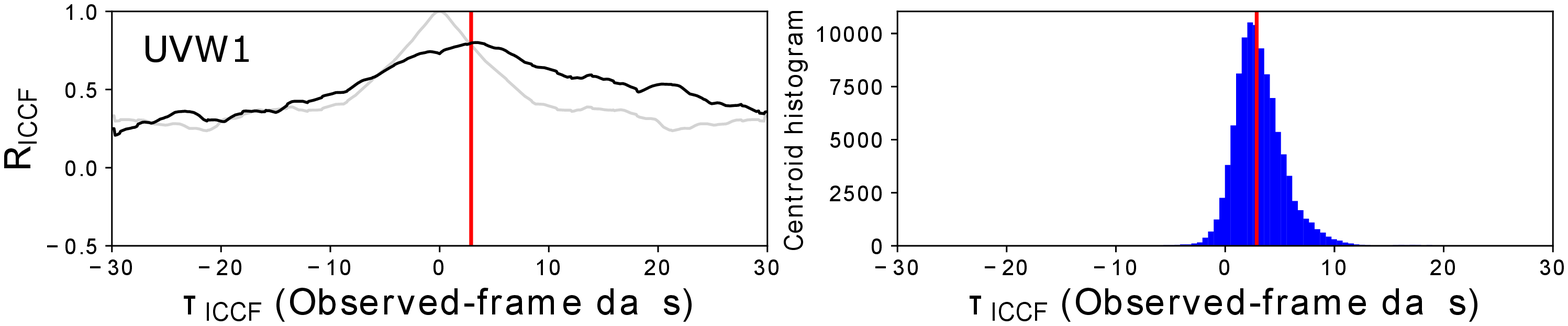}
   \includegraphics[scale=\picscale,trim=0 30 0 0, clip]{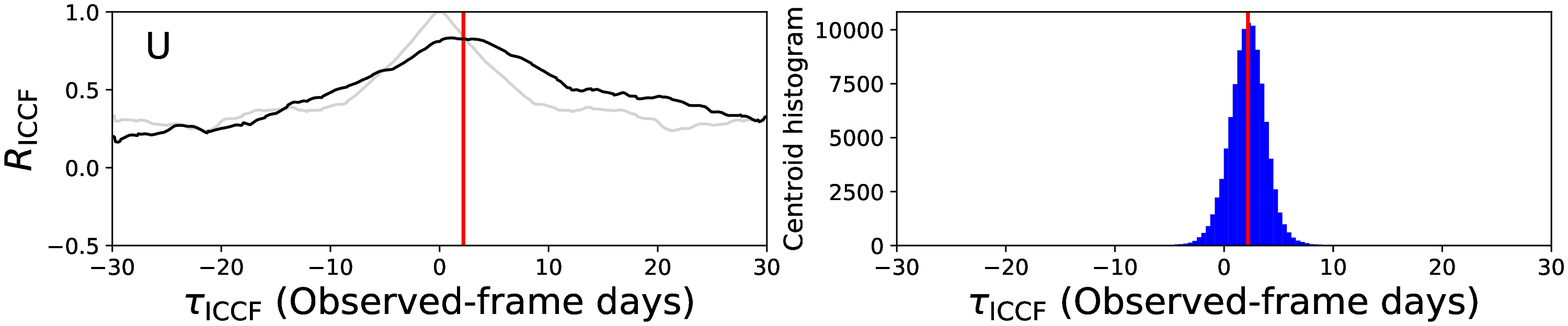}
   \includegraphics[scale=\picscale,trim=0 30 0 0, clip]{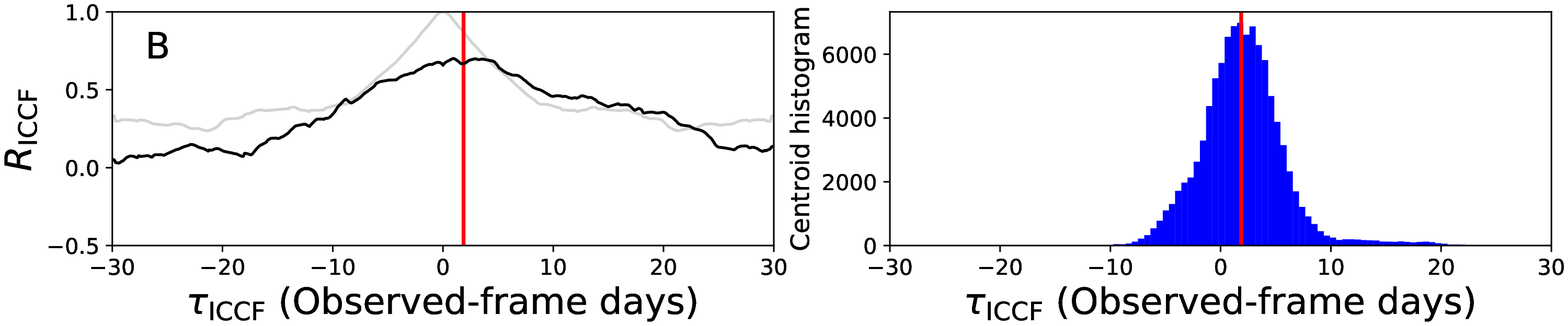}
   \includegraphics[scale=\picscale,trim=0 0 0 0, clip]{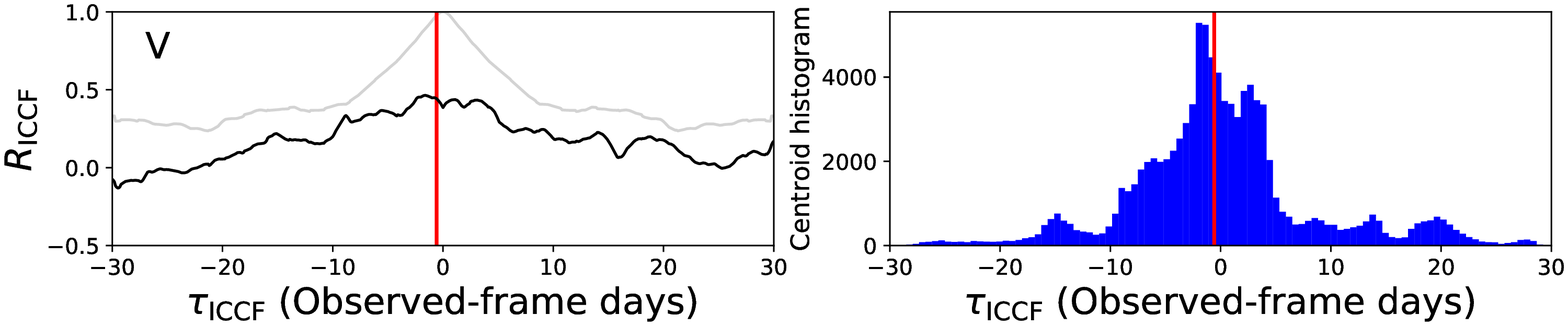}
   \caption{Interpolated Cross-Correlation Function (ICCF) analysis for the XRT 0.3-2 keV (\emph{top panels}), and the UVOT \emph{UVW2} through \emph{V} lightcurves, relative to the XRT 2-10 keV lightcurve. The vertical red lines denote the median centroid for 100,000 Monte Carlo realizations of the lightcurve, with flux randomization and random subset selection. \emph{Left:} the interpolated cross-correlation function (CCF) for the observed data (black curves). For comparison purposes, the autocorrelation function for the 2-10 keV driving lightcurve is shown in each panel (gray curves). \emph{Right:} The CCF centroid distributions for the Monte Carlo realizations with flux randomization and random subset selection.}
   \label{fig:iccf_lags}
\end{figure*}

The interpolated cross-correlation function (ICCF) method \citep{White1994} calculates the cross-correlation function (CCF) between two lightcurves with non-uniform time sampling using linear interpolation between the discrete data points. Here, we use the \textsc{PyCCF} \citep{PyCCF2018} implementation\footnote{URL: \url{http://ascl.net/code/v/1868}} of the ICCF method. In this analysis, we treat the hard (2--10 keV) X-ray lightcurve as the driving lightcurve. For each \emph{Swift} UVOT lightcurve, and for the soft (0.3-2 keV) X-ray lightcurve, we extract the CCF with respect to the driving lightcurve. We calculate the CCF once while interpolating over the driving lightcurve, once more while interpolating over the response lightcurve, and present the average of these two functions as our final CCF. The correlation strength $R_\mathrm{ICCF}$ corresponds to the maximum value of the CCF for the observed data, such that $R_\mathrm{ICCF}=1$ implies that the lightcurves are identical (apart from a constant rescaling factor) when shifted by the corresponding time delay. 

\paragraph*{X-ray to UV--optical correlations:} The hard and soft X-ray lightcurves are strongly correlated, with $R_\mathrm{ICCF}=0.94$ (Table \ref{tab:mrk590_rm}), as suggested by the roughly constant X-ray spectral shape as a function of X-ray flux (Figure \ref{fig:xrt_Gamma}). The X-ray and UV lightcurves are also strongly correlated (Table \ref{tab:mrk590_rm}), e.g., $R_\mathrm{ICCF}=0.87$ for the $UVW2$ lightcurve relative to the hard X-rays. The $B$ lightcurve displays a more modest correlation with the X-ray variability ($R_\mathrm{ICCF}=0.70$), while the $V$ band is only weakly correlated ($R_\mathrm{ICCF}=0.46$).

The strong X-ray to UV correlation (on timescales of a few days) is rather unusual for AGN, at least for the few sources currently studied with continuum reverberation mapping. For example, \citet{Edelson2019} analyze \emph{Swift} monitoring campaigns of four AGN, and find $R_\mathrm{ICCF}<0.7$ for the hard X-ray to $UVW2$ correlation in all cases. We demonstrate in Appendix \ref{sec:appendixA} that our $R_\mathrm{ICCF}$ values are unlikely to be strongly biased by our rather sparse ($\sim2$-day) observational sampling. Based on our simulations, we suggest a lower limit $R_\mathrm{ICCF}>0.75$ for the X-ray to $UVW2$ correlation, \emph{accounting for the use of a $\sim2$-day sampling instead of $\sim0.5$-day}. We discuss these strong correlations in the context of CLAGN activity in \S \ref{sec:discussion_correlation}.

\paragraph*{Time delay measurements:} To determine the time delays $\tau_\mathrm{ICCF}$, and their uncertainties due to flux errors and discrete time sampling, we generate 100,000 Monte Carlo realizations of each lightcurve pair, applying flux randomization and random subset selection \citep{Peterson1998}. For each realization, we determine a Gaussian centroid of the CCF. We then use the distribution of centroids for a given lightcurve pair to determine $\tau_\mathrm{ICCF}$ and its uncertainty. We present the observed CCFs and the Monte Carlo centroid distributions for each lightcurve, relative to the 2--10 keV X-ray lightcurve, in Figure \ref{fig:iccf_lags}. The 0.3--2 keV response is consistent with zero lag. For the UV bands, the centroid distributions peak at between 2.2 and 2.9 observed-frame days. Their centroid distributions are rather broad; all UV lags are consistent with $\sim2.5$ observed-frame days based on the standard deviations (Table \ref{tab:mrk590_rm}). The responses of the $B$ and $V$ bands are weaker, with correspondingly broader centroid distributions, consistent with zero lag. 

\paragraph*{X-ray autocorrelation function:} For comparison purposes, we show the autocorrelation function for the 2--10 keV lightcurve as gray curves in the left panels of Figure \ref{fig:iccf_lags}. The width of the autocorrelation function peak suggests that typical minor flares during the 2017--2018 flare-up have a duration of $\lesssim10$ days. All X-ray to UV cross-correlation functions display broader peaks than that of the autocorrelation function, suggesting reprocessing of the X-ray variations with some additional temporal smoothing, as expected if a compact source (here, the X-ray corona) drives a more extended reprocessor \citep[e.g.,][]{Collier1999}.

\begin{table}
\caption{\textsc{Javelin} and ICCF reverberation mapping results.\label{tab:mrk590_rm}}
\begin{tabular}{cccc}
\hline
\hline
	Bandpass & $\tau_\mathrm{J}$  & $R_\mathrm{ICCF}$ & $\tau_\mathrm{ICCF}$ \\
	 & (days) &  & (days) \\
	(1) & (2) & (3) & (4) \\
	\hline
	XRT 0.3-2 keV & 0.1$_{-0.3}^{+0.1}$ & 0.94 & 0.5$\pm0.7$ \\
	UVOT \emph{UVW2} & 2.8$^{+0.6}_{-0.5}$ & 0.87 & 2.5$^{+1.4}_{-1.6}$ \\
	UVOT \emph{UVM2} & 2.8$\pm0.6$ & 0.86 & 2.6$^{+1.3}_{-1.7}$ \\
	UVOT \emph{UVW1} & 3.3$\pm0.6$ & 0.80 & 2.9$^{+1.8}_{-2.3}$ \\
	UVOT \emph{U} & 2.9$^{+0.6}_{-0.5}$ & 0.83 & 2.2$^{+1.6}_{-1.5}$ \\
	UVOT \emph{B} & 1.7$^{+1.3}_{-1.5}$ & 0.70 & 1.9$^{+3.3}_{-3.1}$ \\
	UVOT \emph{V} & 1.6$^{+3.5}_{-5.5}$& 0.46 & -0.6$^{+6.0}_{-6.2}$ \\
	\hline
	\end{tabular}
	\\

	Columns: (1) \emph{Swift} XRT energy range, or UVOT filter name, for the `response' lightcurve in the analysis. We treat the XRT hard X-ray band (2-10 keV) as the driving lightcurve; all quoted lags are measured relative to 2-10 keV.\\
	(2) Median of \emph{Javelin} posterior distribution for the lag between the XRT 2-10 keV driving lightcurve and this band, in units of observed-frame days. The quoted uncertainties correspond to 90\% Highest Posterior Density intervals as calculated by \emph{Javelin}. \\
	(3) Maximum correlation strength of the interpolated cross-correlation function, for the listed bandpass relative to the XRT 2-10 keV driving lightcurve.\\
	(4) Median time delay for the distribution of ICCF centroids, for 100,000 realizations of the lightcurves including flux randomization and random flux resampling. The quoted uncertainties correspond to the $1\sigma$ width of the centroid distribution.\\
\end{table}

\subsection{\textsc{Javelin} method}\label{sec:javelin}

\begin{figure}
   \newcommand{\pdfscale}{0.265}
   \centering
   \includegraphics[scale=\pdfscale]{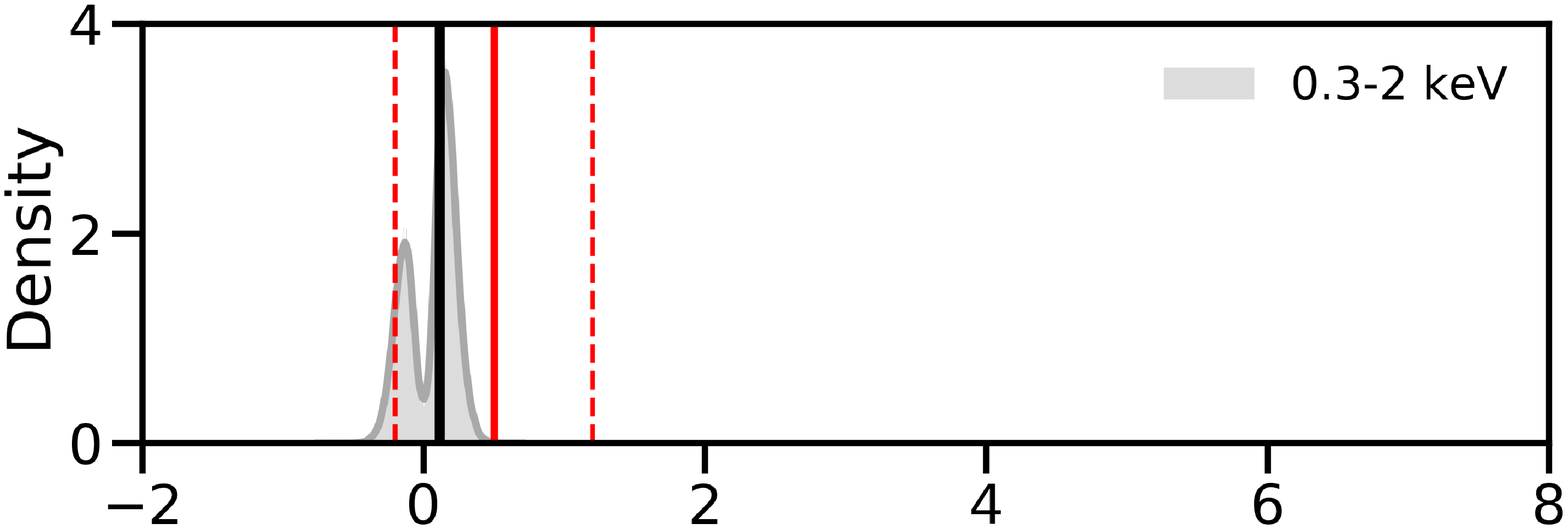}
   \includegraphics[scale=\pdfscale]{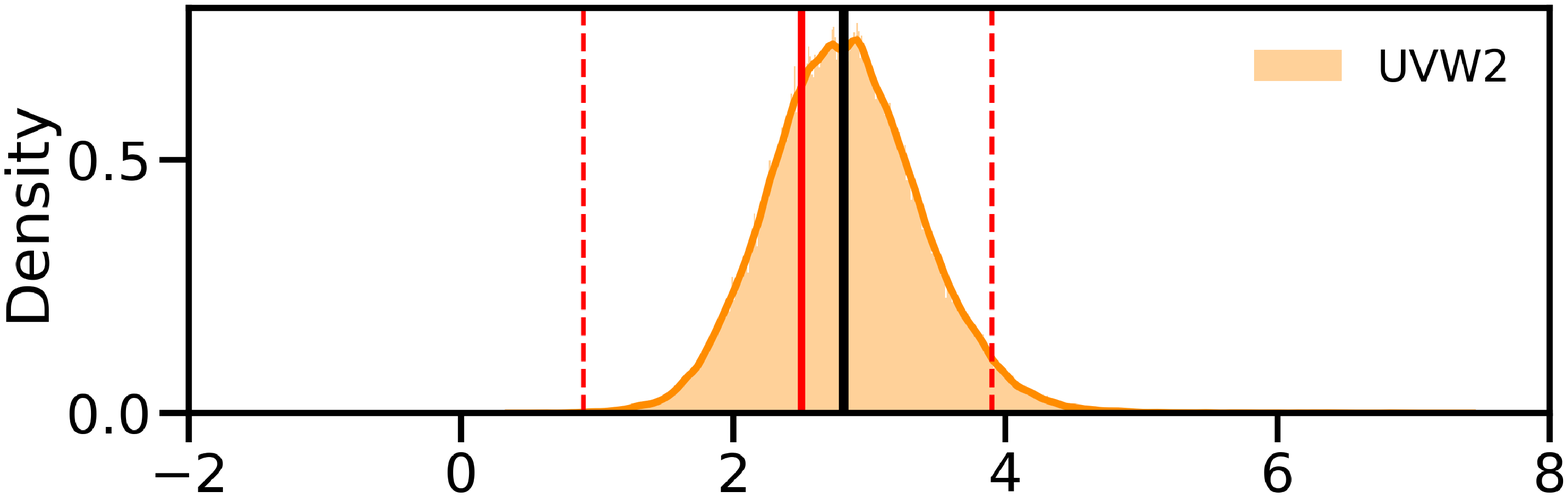}
   \includegraphics[scale=\pdfscale]{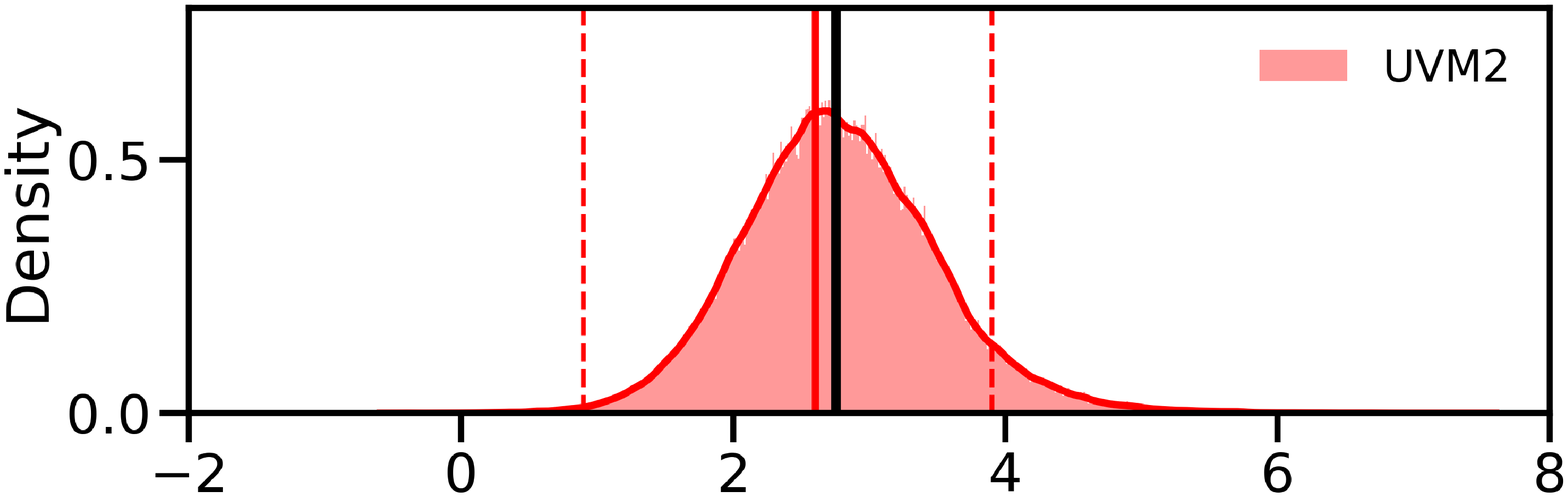}
   \includegraphics[scale=\pdfscale]{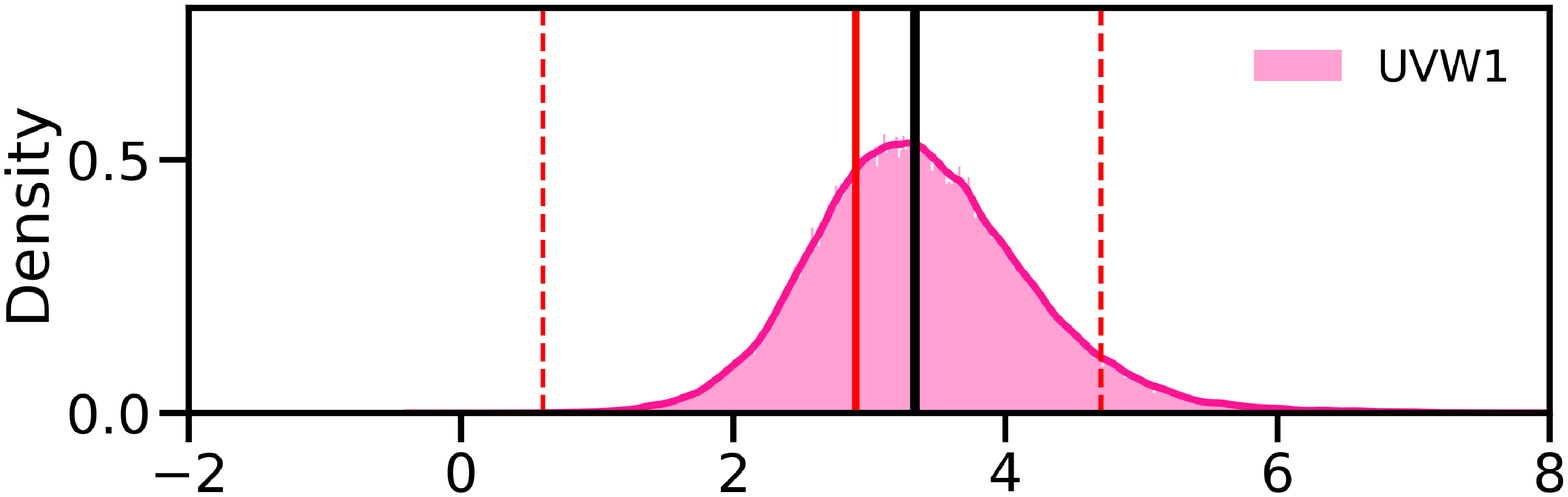}
   \includegraphics[scale=\pdfscale]{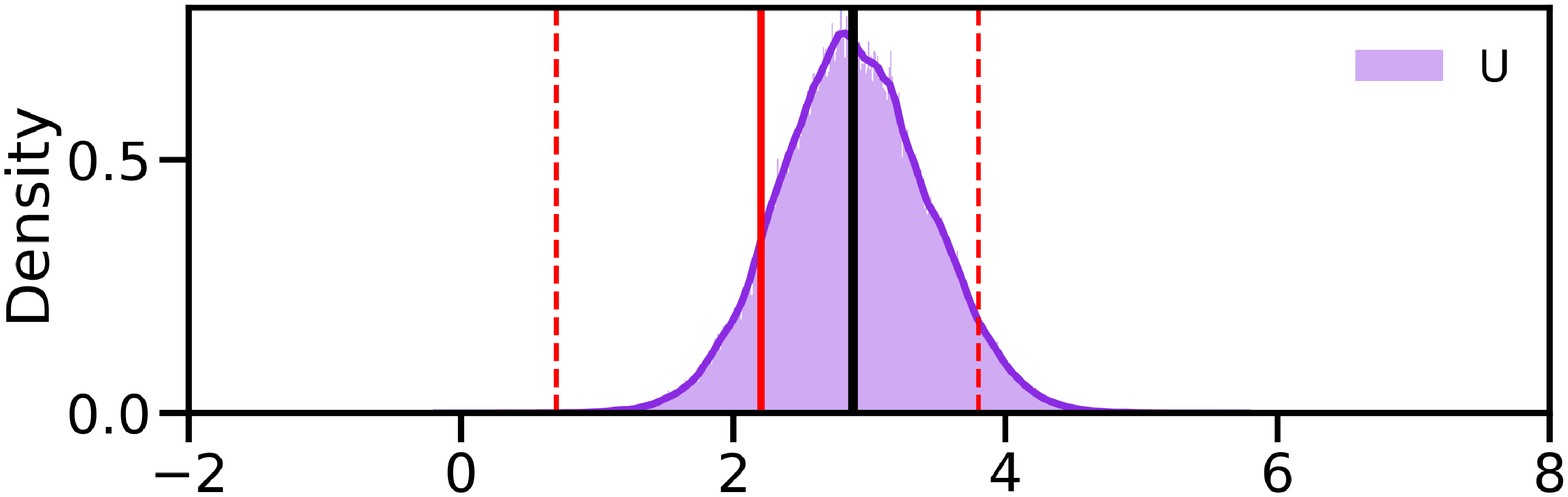}
   \includegraphics[scale=\pdfscale]{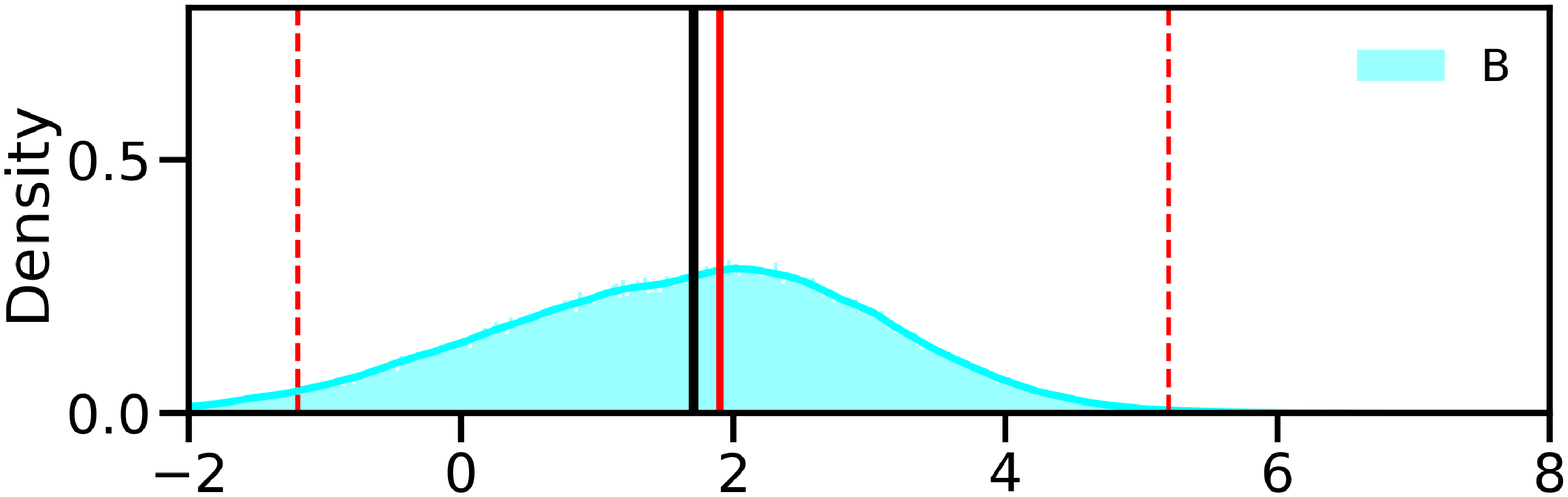}
   \includegraphics[scale=\pdfscale]{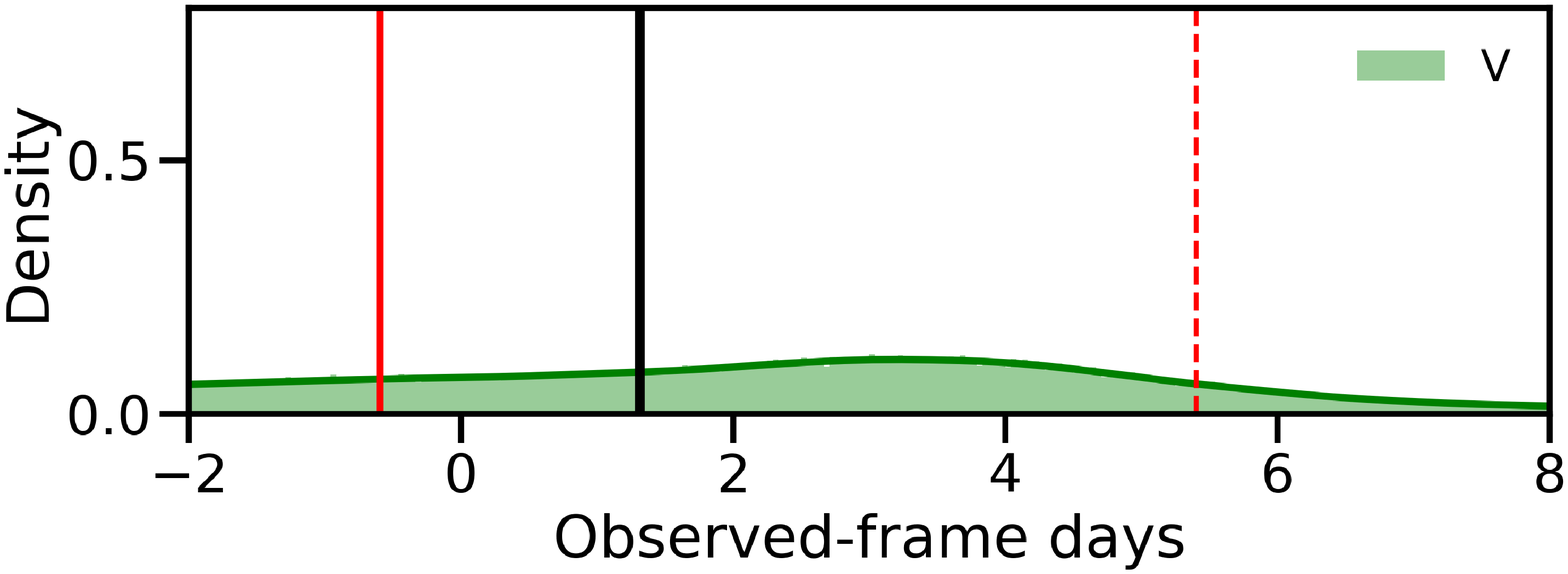}
   \caption{Histograms of \emph{Javelin} Monte Carlo Markov Chain lag distributions for the soft X-ray band (0.3--2 keV), and for each UVOT filter, relative to the XRT 2--10 keV lightcurve, during 8th August 2017 -- 29th February 2018. The colored histograms represent the lag distributions. The black vertical lines display the median \textsc{Javelin} lag for each band relative to the 2--10 keV lightcurve. We tested for lags of -30 to +30 observed-frame days, but find no additional peaks beyond the ranges shown here. The $V$ band lag distribution is single-peaked but very broad, extending beyond the limits of this figure. For comparison, the red solid lines indicate the ICCF median centroid, and the red dashed lines indicate the $1\sigma$ width of the ICCF centroid distribution.}
   \label{fig:javelin_lags}
\end{figure}

The \textsc{Javelin} software \citep{Zu2011} provides an alternative method of estimating reverberation mapping time delays. \textsc{Javelin} explicitly treats one of the lightcurves as a \emph{driving continuum}, and models it as a first-order autoregressive process, as described in \S \ref{sec:results_lc}. \textsc{Javelin} then models the variable component of the `response' lightcurves as shifted, smoothed (via a top-hat transfer function), and rescaled versions of the modeled driving continuum. A Monte Carlo Markov Chain technique is used to find the maximum-likelihood parameters for a joint model consisting of \emph{1)} the model driving continuum, and \emph{2)} the time delays, transfer function widths, scaling factors, and constant-flux components for each reverberating lightcurve. An advantage of this approach is its use of information from all response lightcurves simultaneously to constrain the continuum model, instead of relying on linear interpolation.

We perform a simultaneous \textsc{Javelin} analysis of our XRT 2--10 keV and 0.3--2 keV, and UVOT $UVW2$ through $V$ lightcurves, and extract posterior distributions of model parameters from the Monte Carlo Markov chains. As in our ICCF analysis, we treat the 2--10 keV lightcurve as the driving continuum. The 2--10 keV lightcurve has a damping timescale of $\tau=12.3_{-3.3}^{+4.8}$ observed-frame days, and a fractional variability amplitude of 3.2$_{-0.2}^{+0.6}$. 

\paragraph*{\textsc{Javelin} time delays:} The 0.3--2 keV band displays a sharp, double-peaked posterior lag distribution near zero lag. Given our rather sparse time sampling ($\sim2$ observed-frame days), we do not assign any physical significance to the shape of this distribution, and simply note that it is consistent with zero lag. For the UVOT $UVW2$ through $U$ bands, the posterior lag distributions relative to the 2--10 keV lightcurve are single-peaked, with median values of $\sim3$ observed-frame days (Figure \ref{fig:javelin_lags}). The X-ray to UV delays for the $UVW2$ through $U$ bands are inconsistent with zero lag at the $>3\sigma$ level, based on the 90\% Highest Posterior Density intervals for each parameter (Table \ref{tab:mrk590_rm}). The $B$- and $V$-band lag distributions are broader than those of the UV bands; the lags are not well constrained by the data (Figure \ref{fig:javelin_lags}, bottom two panels). This is expected given the low correlation strength measured for these bands using the ICCF method, indicating a weak response to X-ray variability. While the median \textsc{Javelin} lags are consistent with those of our ICCF analysis, their uncertainties are smaller; the 2--10 keV to UV lag distributions produced by \emph{Javelin} are clearly inconsistent with zero. Simulation studies show that \textsc{Javelin} tends to produce equally accurate lag estimates to the ICCF technique, while yielding more accurate uncertainties \citep{Li2019,Yu2020}. We therefore find it likely that the measured delays are real and non-zero. 

\paragraph*{Choice of driving continuum lightcurve:} While our ICCF analysis treats lightcurve pairs `symmetrically' when calculating the CCF, the \emph{Javelin} analysis explicitly models the driving lightcurve as an AR(1) process. To test this, we repeat the \emph{Javelin} analysis twice, \emph{i)} using the 0.3--2 keV band as the driving lightcurve, and \emph{ii)} using $UVW2$ as the driving lightcurve. In both cases, the results are qualitatively similar to those of our initial analysis: the 0.3--2 keV to 2--10 keV lag is consistent with zero, while the X-ray to UV lags are consistent with $\sim3$ days. We therefore use the lags obtained using the 2--10 keV lightcurve as the driving continuum in the remainder of this work.

\paragraph*{Transfer function widths:} Our \textsc{Javelin} analysis yields rather long posterior-median widths for the top-hat transfer functions, of order 5-10 days, i.e., longer than the measured lags. The transfer function widths can in principle be used to constrain the `lamp-post' geometry introduced in \S \ref{sec:discussion_rm}, as they depend on the X-ray source scale height, due to both geometrical and irradiation effects \citep{Kammoun2021}. However, if the photometric uncertainties for the response lightcurves are large relative to their variability amplitudes, the widths measured by \textsc{Javelin} will tend towards larger values even for a narrow underlying response function (Y. Zu, \emph{priv.comm.}). We find that multiplying the UV--optical photometric error bar sizes by a factor 0.9 yields top-hat widths of $\sim0.5$ days, with median lags fully consistent with our original analysis. This high sensitivity to the error-bar scaling indicates that our data do not warrant inclusion of the width as a free parameter. We repeat our analysis with the transfer function widths held constant (1 observed-frame day), again finding lags fully consistent with those of Figure \ref{fig:javelin_lags}. We conclude that our data do not constrain the transfer function widths, but that the lags are robust.

\subsection{Dependence of measured lags on variability frequency}\label{sec:results_rm_filtering}

\begin{figure}
   \newcommand{\picscale}{0.6}
   \centering
   \includegraphics[scale=\picscale,trim=15 0 0 0, clip]{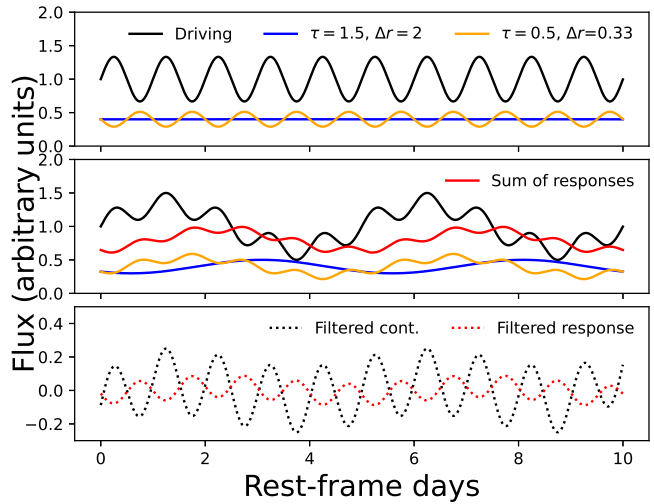}
   \caption{Illustration of the effects of reprocessor size on the response to rapid and slow continuum variability. To emphasize these effects, here we use a sinusoid as a simple `driving continuum' model, use top-hat response functions for the reprocessors, and neglect photometric uncertainties. \emph{Top panel:} A rapidly varying continuum (1-day period, black curve) drives a compact reprocessing region with an emissivity-averaged delay $\tau$=0.5 light-days and a spatial extent $\Delta r=$0.33 light-days (orange curve). This variability pattern does not produce a response in a second, extended reprocessor with $\Delta r=$2 light-days (blue curve).  \emph{Middle:} Here, we add an additional sinusoidal variation to the continuum, with a 5-day period. The compact reprocessor responds to both the high-and low-frequency variability, while the extended reprocessor responds only to the low-frequency variability content. In a real observing situation we would measure the summed lightcurve of the two reprocessor components (red curve). \emph{Bottom:} Applying the `smooth-and-subtract' filtering technique to the continuum and summed response lightcurves shown in the middle panel, we approximately recover the response of \emph{only the compact reprocessor} to the high-frequency variability, demonstrating the applicability of our filtering approach.}
   \label{fig:filtering_illustration}
\end{figure}

\begin{figure*}
   \newcommand{\picscale}{0.53}
   \centering
   \includegraphics[scale=\picscale,trim=0 30 0 0, clip]{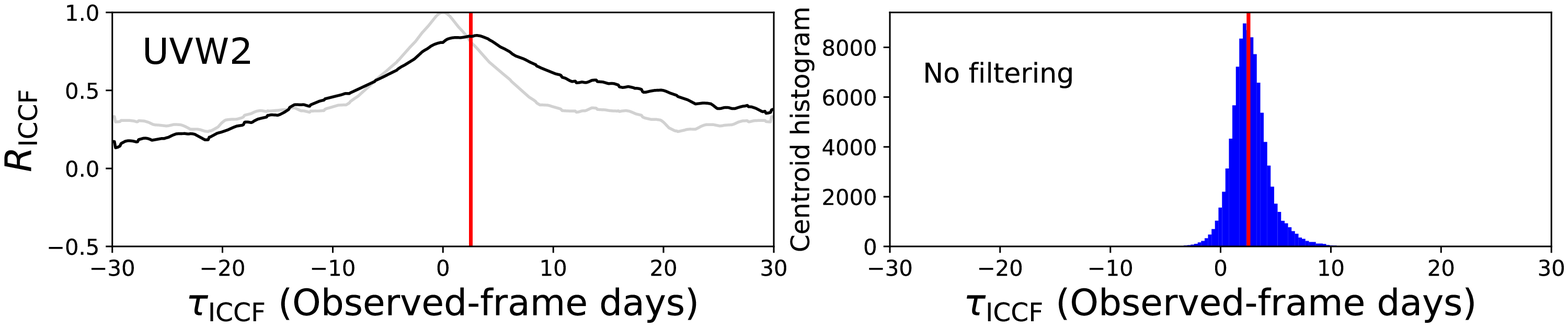}
   \includegraphics[scale=\picscale,trim=0 30 0 0, clip]{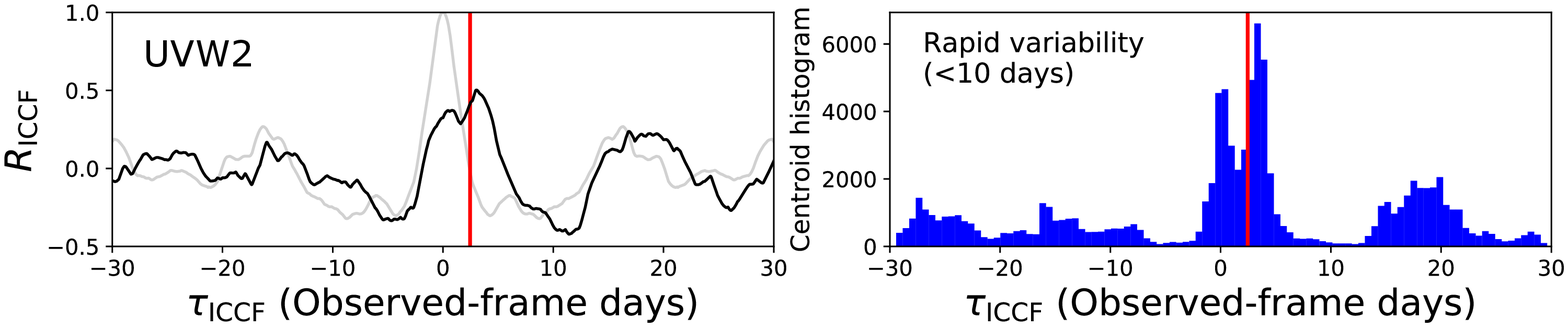}
   \includegraphics[scale=\picscale,trim=0 0 0 0, clip]{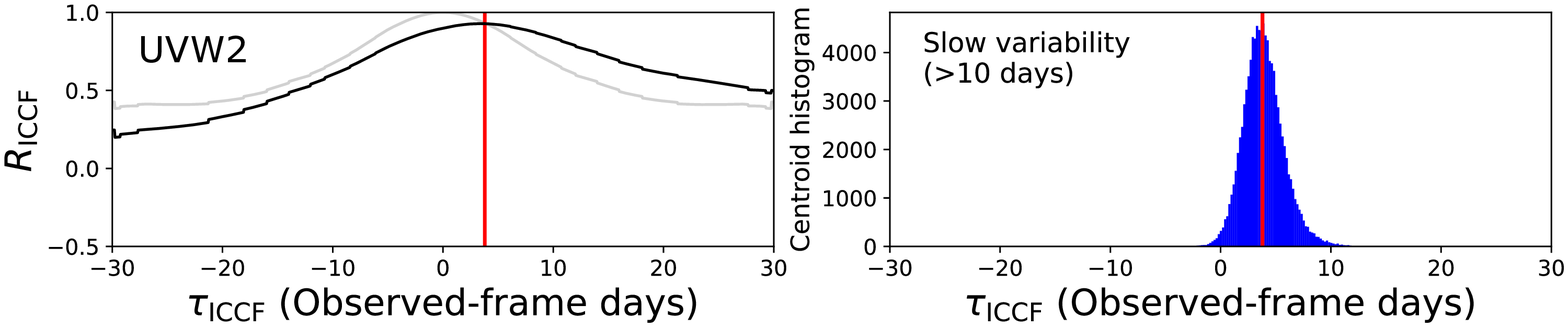}
   \caption{Interpolated Cross-Correlation Function (ICCF) analysis for the UVOT \emph{UVW2} lightcurve, relative to the XRT 2-10 keV lightcurve, after application of the `boxcar smoothing' filtering technique with a smoothing width of 10 observed-frame days. \emph{Top panels:} the cross-correlation function \emph{(left)} and ICCF centroid distribution \emph{(right)} for the unfiltered lightcurves, equivalent to Figure \ref{fig:iccf_lags} and repeated here for comparison purposes. \emph{Middle panels:} the CCF and centroid distribution for the rapid variability, i.e., subtracting the boxcar-smoothed lightcurves (with a 10-day smoothing width) from the observed lightcurves. While the median of the CCF centroid distribution (red vertical line) is similar to that of the unfiltered lightcurves, the distribution itself is double-peaked, as is the CCF itself. \emph{Bottom panels:} the CCF and centroid distribution for the slow variability, i.e., analyzing the boxcar-smoothed lightcurves themselves.}
   \label{fig:iccf_lags_filtered}
\end{figure*}

Several recent disk reverberation mapping studies find evidence for a dependence of the measured inter-band lag on the variability frequency, in the sense that more rapid continuum variations produce a response at a shorter time delay \citep{McHardy2018,Pahari2020,Vincintelli2021,Cackett2021}. These results may be due to additional, spatially extended reprocessing regions situated beyond the accretion disk, which only respond coherently to slow variability. We illustrate this effect using simple sinusoidal continuum models in Figure \ref{fig:filtering_illustration} (top and middle panels). In this Section, we separate the low- and high-frequency variability content in our \emph{Swift} lightcurves, in an effort to `pick out' the reverberation signal of compact reprocessors. While sophisticated, frequency-resolved lag analyses are available \citep{Cackett2021}, our data are rather sparsely sampled and do not warrant this approach. Instead we turn to the `smooth and subtract' technique \citep{McHardy2014,McHardy2018,Pahari2020,Vincintelli2021}, as follows. First, we generate a boxcar-smoothed version of each lightcurve (driving continuum and response), to represent the slow variability. We then subtract the smoothed lightcurve from the observed data to isolate the rapid variability. Figure \ref{fig:filtering_illustration} (bottom panel) illustrates the results of this technique: for an appropriate choice of the boxcar smoothing width, the rapid variability is isolated for our model continuum and response lightcurves. We rely on the ICCF method to analyze the filtered lightcurves. The \textsc{Javelin} approach is not formally applicable here, as there is no guarantee that the rapid- and slow-variability X-ray lightcurves individually correspond to an AR(1) process.

\paragraph*{Choice of boxcar smoothing width:} Boxcar smoothing is equivalent to a $\sinc$-function low-frequency pass filter, with the filter cutoff frequency corresponding to the inverse of the smoothing width. Similarly, \emph{subtracting} a boxcar-smoothed lightcurve from the observed data results in a high-frequency pass filtering. Ideally, given sufficiently strong rapid continuum variability, we would set the boxcar smoothing width to correspond to a few times the expected disk crossing time, in order to isolate the disk response. A too large smoothing width will not filter out the response of more extended reprocessing components. However, if the boxcar width is too narrow \emph{relative to the typical variability timescale of the driving continuum}, the resulting rapid-variability lightcurves become noise-dominated. We experiment with different boxcar smoothing widths for our Mrk 590 lightcurves. A 5-day boxcar width erases most of the variability information, such that no significant lags are recovered in the rapid-variability lightcurves. For the $UVW2$ through $U$ filters, a 10-day smoothing width leads to a detectable lag signal in both the rapid-variability and slow-variability lightcurves, with clear differences between the corresponding CCFs. Using 15- or 20-day smoothing widths also yields a lag signal in both slow- and rapid-variability lightcurves, but the differences are less pronounced. We therefore present results for a 10-day smoothing width here.

\paragraph*{Rapid variability:} We repeat the ICCF analysis using the `smoothed-and-subtracted' lightcurves to analyze the reverberation response to rapid variability. As in our initial analysis (\S \ref{sec:iccf}), we use the XRT 2--10 keV lightcurve as the driving lightcurve. For UVOT \emph{UVW2}, the correlation strength with respect to 2--10 keV is reduced for the rapid variability: we find $R_\mathrm{ICCF}=0.51$, while the unfiltered lightcurves yield $R_\mathrm{ICCF}=0.87$ (Figure \ref{fig:iccf_lags_filtered}, middle left panel). In Appendix \ref{sec:appendixB} we demonstrate that this reduction in correlation strength is attributable to the filtering procedure itself. The simulations in Appendix \ref{sec:appendixB} also reproduce the `noise' in the ICCF centroid distribution at large positive and negative lags (Figure \ref{fig:iccf_lags_filtered}, middle right panel), likely due to the imprint of stochastic features at timescales of $>10$ days in the truncated X-ray auto-correlation function, along with the overall reduction in correlation strength in the filtered data. 

The CCF and centroid distribution for the rapid variability shows a second peak near zero lag (Figure \ref{fig:iccf_lags_filtered}). In Appendix \ref{sec:appendixB} we demonstrate that this second peak is \emph{not} an artifact of the filtering procedure. We note that any correlated uncertainties between the `driving' and response lightcurves during individual observations would also produce a CCF peak at zero lag. These lightcurves are observed by separate instruments (XRT and UVOT), and we are not aware of any instrumental issues that would produce correlated uncertainties. We note that the second peak in the rapid-variability CCF appears rather broad, whereas we would expect a sharp `spike' at zero lag if it were due to correlated errors in individual observations. Also, the second feature peaks at roughly 0.7 light-days in the CCF, while a spurious signal due to correlated errors would peak at exactly zero lag. We therefore find it likely that the second peak is a real, albeit weak, reverberation signal at near-zero lag. We see qualitatively similar results (i.e., a reduced correlation strength and a second CCF peak near zero lag) for the \emph{UM2} and \emph{U} filters. For \emph{UVW1} the correlation strength is very low for the rapid-variability lightcurves ($R_\mathrm{ICCF}=0.39$); in this case we do not see a significant second peak. 

These secondary CCF peaks may indicate that the underlying response function of the UV-emitting region is double-peaked, with a dominant extended reprocessor at a $\sim3$-day lag, and a weaker signal from a compact reprocessor near zero lag. The dominant extended component would then `dilute' the response of the compact component in the unfiltered lightcurves, but responds less coherently in the rapid-variability lightcurves, revealing the compact component. Similar to our results, \citet{Cackett2021} find a strong response at near-zero lags for the rapid variability in their analysis of NGC 5548. However, when isolating the rapid variability in our Mrk 590 lightcurves, we still see the strongest response at lags of $\sim3$ days. This may imply that the spatial extent of this $\sim3$-day lagged reprocessor is of order $\sim10$ light-days, such that the response is still semi-coherent for these variability timescales. Unfortunately, our data do not allow the isolation of even shorter variability timescales with which to test this hypothesis. In future reverberation mapping studies of Mrk 590, it is important to obtain an improved observational cadence in order to probe the most rapid variability behavior.

\paragraph*{Slow variability:} We also perform the ICCF analysis for the boxcar-smoothed lightcurves themselves, in order to isolate the reverberation response to slow continuum variations. For the \emph{UVW2} through \emph{U} filters, we find a higher correlation strength for the slow variability than for the unfiltered lightcurves, e.g., $R_\mathrm{ICCF}=0.93$ for \emph{UVW2} (Figure \ref{fig:iccf_lags_filtered}, bottom left panel). This is in part a natural consequence of filtering out the \emph{uncorrelated} instrumental noise in each observation. For all UV filters, the ICCF centroid distribution is shifted towards longer lags for the slow-variability lightcurves (Figure \ref{fig:iccf_lags_filtered}, bottom panels). For \emph{UVW2} we measure $\tau_\mathrm{ICCF}=3.8^{+1.6}_{-1.8}$ observed-frame days for the slow variability, compared to $\tau_\mathrm{ICCF}=2.8^{+0.6}_{-0.5}$ for the unfiltered lightcurves. While this shift is not significant relative to the standard deviations of the centroid distributions for each individual UVOT filter, it is nonetheless consistent with the `filtering out' of an additional weak reprocessing component at near-zero lag.

\section{Discussion}\label{sec:discussion}

Our \emph{Swift} monitoring observations capture the re-ignition of Mrk 590 from a low-luminosity state during August 2017, and its subsequent repeat X-ray and UV flaring behavior. The main results presented in this work are:

\begin{itemize}
    \item We observe strong variability in the X-rays and UV since 2017, with characteristic timescales $\tau_\mathrm{char}\sim100$ days
    \item The X-ray and UV lightcurves are highly correlated
    \item The UV lightcurves lag the X-rays by $\sim3$ rest-frame days. We do not detect any UV--optical inter-band lags within our temporal sensitivity of $\sim1.5$ days, due to the average cadence of the monitoring data (\S \ref{sec:discussion_rm})
\end{itemize}

We now discuss each of these results in turn. 

\subsection{The UV and X-ray flaring activity}\label{sec:discussion_flares} 

\subsubsection*{The reappearance of the UV continuum:} Typical broad-line `Type 1' AGN display a blue UV--optical continuum, along with both broad and narrow emission lines. During the 2013 low-luminosity state, Mrk 590 lost its broad Balmer emission lines and optical AGN continuum emission. At the same time, its UV continuum and broad Ly$\alpha$ and C \textsc{iv} emission lines became vary faint, essentially appearing as a `Type 2' AGN \citep{Denney2014}. During 2017, Mrk 590 displays an abrupt increase in both X-ray and UV luminosity, and in variability (\S \ref{sec:results_lc}). The average X-ray flux since August 2017 is $\langle F_\mathrm{0.3-10}\rangle=1.5\times10^{-11}$ erg s$^{-1}$ cm$^{-2}$, a \mbox{factor $\sim5$} higher than the 2014 low state. Similarly, the average far-UV luminosity increases by a factor $\sim6$ compared to the low state. The H$\beta$ broad emission line had reappeared by September 2017 \citep{Raimundo2019}, during the first major flare-up observed by \emph{Swift}. The Mg \textsc{ii} emission line was already present at lower continuum luminosities in 2014 \citep{Mathur2018}. Thus, Mrk 590 transitioned back into a `Type 1' state (at least with respect to its broad emission lines) as the X-ray and UV flux increased, at some point during 2014--2017. Broad Mg \textsc{ii} may never have fully disappeared: the relevant wavelengths were not observed during the 2013 low-luminosity state during which the changing-look behavior was first discovered. As the Mg \textsc{ii} line tends to respond weakly to ionizing continuum variability in other AGN \citep{Cackett2015}, even for extreme decreases in UV continuum luminosity \citep{Ross2018}, it is plausible that broad Mg \textsc{ii}  was present during 2013.

While our \emph{Swift} UVOT photometric data cannot unambiguously separate continuum and emission line variability, we nevertheless find it very likely that the observed UV flare-ups represent a partial re-ignition of the AGN UV continuum, for the following reasons. Firstly, while the UVOT \emph{UVW2} and \emph{UVW1} bandpasses do sample the C \textsc{iii]} and Mg \textsc{ii} emission lines, respectively, the UVOT \emph{U} band does not sample any prominent broad emission lines. However, the \emph{U} band displays a similar abrupt increase in flux and variability during 2017, with major flares coinciding with those observed in the far--UV and X-rays. The \emph{U} band does sample diffuse continuum emission from the broad line region; we explore this possibility in \S \ref{sec:discussion_rm}, but find that diffuse continuum is unlikely to dominate the UV emission. Secondly, an extreme--UV continuum is in any case required to provide ionizing radiation to produce the broad Balmer lines observed in 2017. Finally, our variability analysis (\S \ref{sec:results_lc}) indicates that the variable component is very blue, as expected for the power-law continuum typical of Type 1 AGN activity. The quasi-simultaneous appearance of flares in the X-rays and UV confirm that the excess UV emission is related to the central engine, and not to other variable processes in the host galaxy. Thus, it seems reasonable to attribute the UV flares to broad-band continuum emission from the central source, although it is unclear whether this emission occurs in a thin accretion disk, as discussed further in \S \ref{sec:discussion_rm}.

\subsubsection*{High variability at a modest accretion rate:} Relative to other $z\lesssim0.1$ AGN with well-sampled lightcurves on timescales of months to years, the excess fractional variance of Mrk 590 since 2017 is rather high, but not unprecedented. The UV lightcurves display $F_\mathrm{var}\sim30$\% after the initial flare-up. For other AGN monitored by \emph{Swift}, $F_\mathrm{var}$ is typically 4\%--23\% in the \emph{UVW2} filter \citep[e.g.,][]{Gallo2018,Edelson2019,Lobban2020,Cackett2020,Hernandez2020}. In the X-rays, we find $F_\mathrm{var}$=47\% for Mrk 590 since August 2017, compared to typical values of $\sim10$--35\% for moderately accreting AGN \citep{Edelson2019,Lobban2020,Hernandez2020,Kumari2021,Vincintelli2021}. In fact, the X-ray variability amplitude of Mrk 590 is similar to that of Narrow Line Seyfert 1 sources, which in many cases display $F_\mathrm{var}\sim$50\% \citep[e.g.,][]{Gallo2018,Cackett2020,Ding2022}. Narrow Line Seyfert 1 sources tend to have high accretion rates, near or above the Eddington limit \citep[e.g.,][]{Jin2012}. In contrast to this, Mrk 590 does not currently appear to be highly accreting. Based on spectral energy distribution modeling, we estimate that its post-2017 accretion rate is typically around $\sim2$\% and reaches a maximum of only $\sim5$\% of the Eddington ratio (\emph{Lawther et al., in prep.}). Since 2017, Mrk 590 is thus highly X-ray and UV-variable at a low Eddington ratio. We speculate that this is related to its recent changing-look events. CLAGN tend to display lower Eddington ratios than `steady-state' AGN \citep{MacLeod2019}, as do extreme-variability AGN more generally \citep{Rumbaugh2018}. 

\subsubsection*{Variability on the disk thermal timescale?} We find characteristic timescales of $\tau_\mathrm{char}\sim100$ rest-frame days for the \emph{UVW2} through \emph{B} bands (\S \ref{sec:results_lc}); the \emph{V} band is not sufficiently variable to yield a $\tau_\mathrm{char}$ estimate. For the X-rays, we also find $\tau_\mathrm{char}\sim100$ days in our structure function analysis, although the \textsc{Javelin} analysis also indicates a rapidly varying component not seen in the UV. Given the strong correlation between the lightcurves, it is unsurprising that they have similar $\tau_\mathrm{char}$, which likely corresponds to some physical timescale for the driving continuum. To investigate this, we calculate the physically relevant timescales for a standard \citet{Shakura1973} `thin disk' model. We assume an accretion disk with an Eddington accretion ratio $\dot{M}_\mathrm{Edd}=$5\%, and a black hole mass $M_{\mathrm{BH}}=3.7(\pm{0.6})\times10^{7}$ $M_{\astrosun}$  \citep{Bentz2015}. For this standard thin-disk model, the radius $r$ at which the disk emits at a characteristic wavelength $\lambda$ is given by:

\begin{figure}
    \centering
    \includegraphics[scale=0.55]{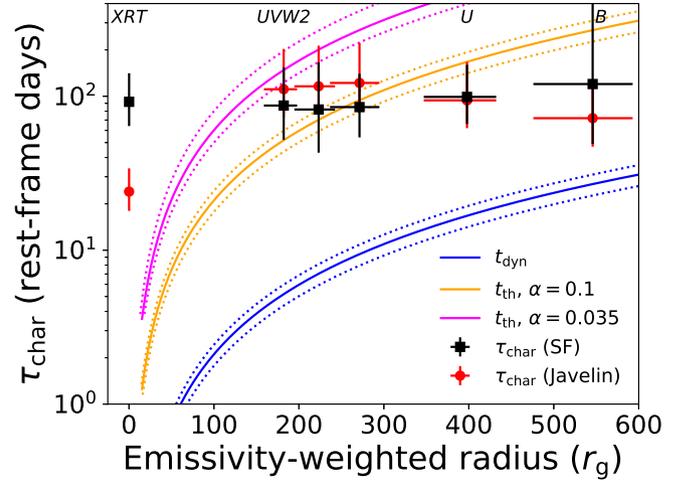}
    \caption{Comparison of the dynamic (\emph{solid blue curve}) and thermal (\emph{solid yellow curve}) timescales, as a function of radius from the black hole, with the characteristic timescales derived from our structure function analysis (\emph{black squares}) and \textsc{Javelin} modeling (\emph{red circles}). We do not show the viscous timescale here; it is of order $10^5$ years for Mrk 590. The dotted blue and yellow lines represent the $1\sigma$ lower and upper limits on $t_\mathrm{th}$ and $t_\mathrm{dyn}$, only accounting for the statistical uncertainty on the black hole mass, $M_\mathrm{BH}=3.7(\pm{0.6})\times10^{7}M_{\astrosun}$ \citep{Bentz2015}. The horizontal error-bars on our $\tau_\mathrm{char}$ estimates represent the uncertainty on the emissivity-weighted radius for the central wavelength of the bandpass (Equation \ref{eq:lamppost}), again accounting only for the $M_\mathrm{BH}$ uncertainties. All radii are calculated using Equation \ref{eq:lamppost}, and expressed in units of the gravitational radius, $r_\mathrm{g}=GM_\mathrm{BH}/c^2$. We assume an Eddington accretion ratio $\dot{M}_\mathrm{Edd}=0.05$. The thermal timescale is inversely proportional with the viscosity parameter, $\alpha$, which is not well constrained for individual AGN. The measured far-UV $\tau_\mathrm{char}$ remains consistent with the thermal timescale for $\alpha\gtrsim0.035$ (\emph{magenta curve}).}
    \label{fig:timescales}
\end{figure}

\begin{equation}\label{eq:lamppost}
    r=0.09\left(X\frac{\lambda}{1928\mathrm{\text{\AA}}}\right)^{4/3}M_8^{2/3}\left(\frac{\dot{M}_{\mathrm{Edd}}}{0.10}\right)^{1/3}\mathrm{\,\,light-days}
\end{equation}

\noindent\citep[e.g.,][]{Cackett2007,Edelson2017}. Here, $M_8$ denotes the black hole mass in units of $10^8$ $M_\odot$, and $X$ is a scaling factor describing the temperature distribution in the disk. For the \citet{Shakura1973} solution, the local blackbody temperature scales as $T\propto R^{3/4}$, in which case a flux-weighted calculation yields $X=2.49$ \citep{Edelson2017}. To calculate the relevant timescales, we use the approximations provided by \citet{Noda2018}. The dynamic (i.e., orbital) timescale at radius $r$ is given by:

\begin{equation*}
    t_\mathrm{dyn}\approx\sqrt{\frac{GM_\mathrm{BH}}{r^3}}.
\end{equation*}

\noindent For Mrk 590, $t_{\mathrm{dyn}}$ is of order only $\sim5$ days in the UV-emitting inner disk. Local thermal dissipation in a thin disk occurs on the thermal timescale, which is given by:

\begin{equation*}
    t_\mathrm{th}\approx\frac{t_\mathrm{dyn}}{\alpha}.
\end{equation*}

\noindent Here, $\alpha$ denotes the viscosity parameter of the thin-disk model. Simulation studies for thin accretion disks yield $0.02\lesssim\alpha\lesssim0.1$ \citep[e.g.,][]{ONeill2009,Parkin2013,Mishra2016}. For Mrk 590, this suggests $t_{\mathrm{th}}\sim10^2$ days in the disk. Finally, the accreting gas drifts inwards on the viscous timescale:

\begin{equation*}
    t_\mathrm{vis}\approx\frac{t_\mathrm{dyn}}{\alpha}\left(\frac{H}{R}\right)^{-2},
\end{equation*}

\noindent where $H/R$ is the ratio of disk scale height to radius. For a thin disk, $H/R$ is small, and the viscous timescales are then of order $\sim10^5$ years. While various effects such as radiation pressure likely serve to increase $H/R$ in the inner disk of real AGN \citep[e.g.,][]{Dexter2019}, it would require $H/R\approx1$ (i.e., well outside the `thin disk' regime) to reconcile the observed $\tau_{\mathrm{char}}$ with the viscous timescale.

Our measured $t_{\mathrm{char}}$ are thus consistent with the thermal timescale in the UV-emitting inner disk for $0.035\lesssim\alpha\lesssim0.1$ and are inconsistent with the dynamic or viscous timescales \emph{for thin disks} (Figure \ref{fig:timescales}). Mrk 590 is not unusual in this regard: several previous studies find $\tau_\mathrm{char}\sim t_\mathrm{th}$ \citep[e.g.,][]{Collier2001,Liu2008,Gallo2018,Noda2018}. For a sample of 67 AGN spanning a mass range $10^4M_{\astrosun}<M_\mathrm{BH}<10^{10}M_{\astrosun}$, \citet{Burke2021} find a correlation between $\tau_\mathrm{char}$ and $M_\mathrm{BH}$, which they suggest is due to the dependence of $t_\mathrm{th}$ on $M_\mathrm{BH}$. For Mrk 590, their best-fit correlation predicts $\tau=73^{+11}_{-10}$ days, which is fully consistent with our UV--optical timescales as derived from \textsc{Javelin} modeling. However, we note that the UV--optical variations clearly lag the X-ray variations in Mrk 590 (\S \ref{sec:results_rm}), which complicates any interpretation in terms of accretion disk timescales. While processes on the UV thermal timescale may indeed govern the observed variability, it is unclear why this would lead to an X-ray flare which then takes $\sim3$ days to propagate to the far-UV. We discuss the surprising X-ray to UV delay further in \S \ref{sec:discussion_rm}.

Alternatively, \citet{Sniegowska2020} suggest that CLAGN have a truncated disk with a `puffed up', advective inner region. In their scenario, instabilities occurring on the viscous timescale in the unstable transition region could propagate out into the UV-emitting disk, leading to strong variability at shorter timescales than for a `pure' thin disk. \citet{Pan2021} suggest that magnetically driven winds may govern the timescale of such an instability. Using a black hole mass of $10^7M_{\astrosun}$ in their simulations, similar to that of Mrk 590, they see recurring flares on timescales of months to years.  Similarly, \citet{Ross2018} suggest that changing magnetic fields in the inner disk may provoke changes in the inner-disk accretion flow, leading to changing-look events. We speculate that these disk instability mechanisms may be relevant to the observed repeat flaring behavior in Mrk 590.

\subsubsection*{The soft X-ray excess as a potential driving continuum:} While our measured $\tau_\mathrm{char}$ values are broadly consistent with the UV thermal timescale, we find that the X-ray emission leads the UV variability. This may be a clue that the X-ray variability is somehow governed by thermal processes in the inner accretion disk. For example, the prominent soft X-ray excess observed below $\sim2$ keV in many AGN is suggested to be due to Compton upscattering of UV seed photons. This upscattering requires a `warm corona' near the UV-emitting region, which may occur in the disk atmosphere itself \citep[e.g.,][]{Czerny1987,Magdziarz1998,Done2012}. An origin of the soft excess in the disk surface would then explain the relevance of the UV thermal timescale. Previous studies indeed find empirical correlations between the UV and soft X-ray luminosities for individual sources \citep{Mehdipour2011} and for statistical samples of AGN \citep{Atlee2009}, supporting a connection between the soft excess and UV-emitting disk.

In Mrk 590, the soft excess disappeared at some time between 2006--2011, as the AGN approached its historic low-luminosity state in 2013 \citep{Rivers2012}. It reappeared at a low level as early as 2014 \citep{Mathur2018}. While some soft excess contribution is therefore likely present during our monitoring campaign, we see no indication that it is driving the X-ray variability. If the soft excess component is strongly variable while the X-ray continuum is constant, we would expect a `softer-when-brighter' trend for the X-ray spectral slope across the entire dynamic range. We only find a modest dependence of the X-ray spectral shape on flux (\S \ref{sec:results_lc}, Figure \ref{fig:xrt_Gamma}), with hints of a `U-shaped' trend instead of an unambiguous `softer-when-brighter' behavior. In particular, for our 2017-2018 data used for reverberation mapping, the X-ray flux did not exceed $F_{\mathrm{0.3-10}}=2.5\times10^{-11}$ erg cm$^{-2}$ s$^{-1}$; the most significant spectral softening occurs above this flux level. This indicates that the power-law continuum itself is highly variable during the initial flare-up. The soft excess may nevertheless provide a source of seed photons for the power-law component, and thus ultimately govern the variability \citep[e.g.,][]{Gliozzi2013,Porquet2021}. We will present an analysis of the soft excess variability in Mrk 590, based on archival and on-going observations, in future work.

\subsection{The strong X-ray to UV correlation}\label{sec:discussion_correlation}

\subsubsection*{Comparison with non-CLAGN:} Only a handful of AGN have intensive reverberation mapping observations to date. Compared to the currently available studies, the strong correlation between the X-ray and UV lightcurves for Mrk 590 during 2017--2018 is rather unusual. For non-CLAGN, the correlations between X-ray and far-UV lightcurves tend to be much weaker than the inter-band UV--optical correlations. \citet{Edelson2019} summarize disk reverberation experiments using \emph{Swift} XRT and UVOT for four AGN. For the Seyfert galaxy Mrk 509, these authors do find an X-ray to UV correlation strength $R_\mathrm{ICCF}=0.77$, however in that case the X-rays lag the UV instead of vice-versa. For the three other AGN in their sample, the UV bandpasses appear to lag the X-rays, but all with correlation strengths $R_\mathrm{ICCF}<0.75$. They find stronger inter-band UV--optical correlations, with $R_\mathrm{ICCF}\geq0.85$ between the $UVW2$ and $B$ lightcurves for all four AGN. This general pattern of weak X-ray to UV correlations with $R_\mathrm{ICCF}<0.6$, but stronger UV--optical correlations, is also reported for the AGN Ark 120 \citep{Lobban2020}, Mrk 817 \citep{Kara2021}, Mrk 142 \citet{Cackett2020}, and Fairall 9 \citep{Hernandez2020}

\subsubsection*{Comparison with other CLAGN:}
For CLAGN, stronger X-ray to UV correlations have at times been observed. Intensive reverberation mapping data for CLAGN are often collected upon detection of a major flare, as is also the case for our 2017--2018 observations of Mrk 590. Thus, they may provide a more appropriate comparison, as the UV bands respond to a substantial increase in the luminosity of the driving continuum. \citet{Shappee2014} capture NGC 2617 in an X-ray and UV outburst in 2013, associated with the appearance of broad UV--optical emission lines. This source continued to flare up until at least 2017 \citep{Oknyansky2017}. During 2013--2014, the X-ray to UV correlation strength is $R_\mathrm{ICCF}\sim0.8$, which reduces to $R_\mathrm{ICCF}\sim0.6$ by 2016 \citep{Oknyansky2017}. The lag between the X-ray and far-UV lightcurves is $\sim2$ days, as determined using \textsc{Javelin}. Thus, the outburst in NGC 2617 is qualitatively similar to the 2017 re-ignition of Mrk 590, with a strong, delayed UV response to repeated X-ray flares. \citet{Oknyansky2021} present intensive reverberation mapping data for the CLAGN NGC 3516 during a flare-up event in 2020. They find a strong correlation ($R_\mathrm{ICCF}=0.87$) between the X-ray and far-UV lightcurves during the initial February 2020 flare-up, but report that this correlation weakens as the flare dissipates. Conversely, for the CLAGN Mrk 1018, \citet{Lyu2021} find low values of $R_\mathrm{ICCF}\sim0.4$ between the UV and X-ray lightcurves. However, these reverberation mapping observations were performed while Mrk 1018 was in a low-luminosity state. Different accretion physics are likely relevant during the Mrk 1018 low-state observations, relative to the early-onset outbursts in Mrk 590, NGC 2617, and NGC 3516. We speculate that CLAGN may display stronger X-ray to UV correlations \emph{during X-ray outbursts} than non-CLAGN sources, perhaps due to fewer emission components that serve to `smooth out' the temporal response. More disk reverberation measurements for CLAGN are needed in order to confirm any such trend.

\subsection{The 3-day X-ray to UV delay}\label{sec:discussion_rm} 

\begin{figure}
    \centering
    \includegraphics[scale=0.55]{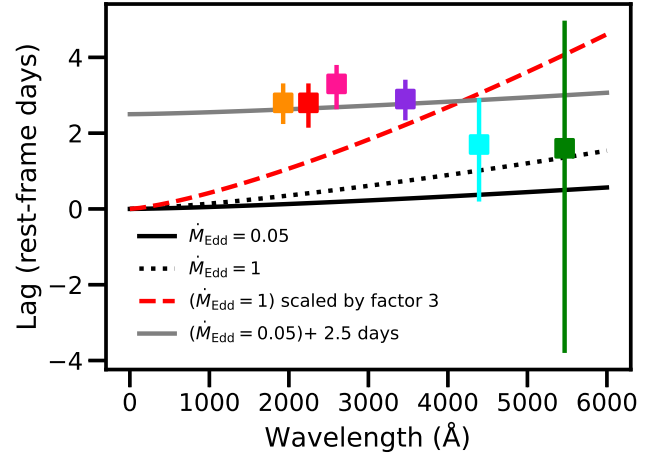}
    \caption{The lag predictions of the `lamp-post' model (\S \ref{sec:discussion_rm}), as a function of wavelength, for a \citet{Shakura1973} disk with $\dot{M}_\mathrm{Edd}=0.05$ (solid black curve), and $\dot{M}_\mathrm{Edd}=1$ (dotted black curve). We compare to the measured \textsc{Javelin} lags for our \emph{Swift} UVOT $UVW2$ (orange square), $UM2$ (red), $UVW1$ (magenta), $U$ (purple), $B$ (light blue), and $V$ (green) lightcurves, relative to the 2--10 keV X-ray lightcurve. By placing the theoretical model curves at $r=0$ for $\lambda=0$, we are implicitly assuming that the X-ray emission occurs very close to the central black hole; the observed X-ray to UV lags are highly inconsistent with this scenario, even assuming accretion at the Eddington limit. Several authors report AGN accretion disk sizes a factor $\sim2$--3 larger than those predicted by the `lamp-post' model. As a crude test for `too-large' disk sizes, we multiply the $\dot{M}_\mathrm{Edd}=1$ model by a factor 3 (red dashed curve); the resulting lag spectrum still underpredicts the far-UV lags. The solid grey curve corresponds to the $\dot{M}_\mathrm{Edd}=0.05$ model curve plus an arbitrary shift of 2.5 light-days. This illustrates that the near-zero UV--optical inter-band lags observed (given the large uncertainties on the $B$ and $V$ band lags) are broadly consistent with the theoretical predictions for $\dot{M}_\mathrm{Edd}=0.05$.}
    \label{fig:lamppost}
\end{figure}

\subsubsection*{Expected delays for a thermal disk:}

For a \citet{Shakura1973} thermal disk illuminated by a compact, variable X-ray source, the so-called `lamp-post' model \citep[e.g.,][]{Cackett2007,Edelson2017} posits that the UV--optical variability is driven by X-ray heating. If the X-ray source is located very near the central black hole (i.e., at a small scale height above the disk), the light travel time to the X-ray source corresponds to the disk radius. In that case, the radial distance at which the disk emits reprocessed X-rays at a characteristic wavelength $\lambda$ is given by \mbox{Equation \ref{eq:lamppost}}. This model predicts that the UV--optical bands are highly correlated with the XRT lightcurve, as is indeed the case for the far-UV and $U$ bands; the weak correlations in \emph{B} and \emph{V} are likely due to host galaxy dilution. The predicted delays for this model depend on the accretion rate, with a higher $\dot{M}$ producing longer delays and a steeper lag spectrum. We estimate that the Eddington-normalized accretion rate is $0.01<\dot{M}_\mathrm{Edd}<0.05$ since 2017, based on our spectral energy distribution analyses (Lawther et al., \emph{in prep.}). In the following we adopt $\dot{M}_\mathrm{Edd}=0.05$; as demonstrated below our findings are not dependent on the precise value of $\dot{M}_\mathrm{Edd}=0.05$ for sub-Eddington accretion.

For our unfiltered lightcurves, the $\sim3$-day lags between the X-ray and far-UV through $B$ bands are significantly longer than the model predictions (\mbox{Figure \ref{fig:lamppost},} solid black curve). Even assuming Eddington-limited accretion ($\dot{M}_\mathrm{Edd}=1$), the lamp-post model alone cannot reproduce the observed X-ray to UV lags (Figure \ref{fig:lamppost}, dotted black curve). An additional, arbitrary $\sim3$-day delay between the X-rays and $UVW2$ is required to reconcile the model and data (Figure \ref{fig:lamppost}, solid gray curve).

We do not detect any UV--optical inter-band lags. This result is not surprising given the $\sim$2-day average time sampling of our monitoring data during 2017-2018. For NGC 5548 and NGC 4151, which have similar black hole masses to Mrk 590, the measured inter-band lags between \emph{Swift} UVOT \emph{UVW2} and \emph{B} filters are 1--1.5 rest-frame days \citet{Edelson2019}. As demonstrated by Figure \ref{fig:lamppost}, the theoretical predictions for the thermal disk are $\sim1$ rest-frame day or less, even considering the extreme case of $\dot{M}_\mathrm{Edd}=1$. Our analysis is insensitive to lags of less than $\sim1.5$ days, as evidenced by the widths of the \textsc{Javelin} posterior distributions (Figure \ref{fig:javelin_lags}). Thus, our monitoring observations lack the sensitivity to detect the expected inter-band lags.

In the following, we explore some possible causes of the unexpected $\sim3$-day delay in Mrk 590.

\subsubsection*{Larger than expected disks?}

Several UV--optical continuum reverberation mapping studies find accretion disk sizes a factor $\sim2$--$3$ larger than those predicted by Equation \ref{eq:lamppost} using the flux-weighted distribution ($X=2.49$) \citep[e.g.,][]{Edelson2019,Ting2021,Montano2022}. Microlensing analyses also suggest larger disk sizes at a given wavelength than predicted by thin-disk models \citep[e.g.,][ and references therein]{Cornachione2020}. To test whether a modest increase in accretion disk size can explain the observed $\sim3$-day X-ray to UV lag, we rescale the $\dot{M}_\mathrm{Edd}=1$ model lag spectrum by a factor 3 (Figure \ref{fig:lamppost}, red dashed curve). This rescaled `lamp-post' model prediction is much steeper than the observed UV--optical lag spectrum, which is rather flat. Also, even assuming a factor 3 increase in disk size, the rescaled `lamp-post' model still under-predicts the X-ray to far-UV lag. Essentially, our non-detection of inter-band UV--optical lags is difficult to reconcile with a scenario where the UV--optical disk itself is much larger (and thus, the lag spectrum much steeper) than the `thin-disk' predictions. 

\subsubsection*{Broad-line reverberation?}

Given our lack of far-UV spectroscopic reverberation mapping data, we need to consider whether the observed response is that of the broad emission lines, and not the UV--optical continuum. If the continuum emission in the UVOT filters is relatively faint during these flares, while the extreme-UV continuum (that powers the broad emission lines) is bright, the broad emission lines may dominate the UV SED. We disfavor this interpretation due to the flat shape of the observed lag spectrum (Figure \ref{fig:lamppost}). In particular, for the source redshift $z=0.02609$, the \emph{Swift} \emph{UM2} filter is largely free of broad emission lines. If the observed response was due to broad line emission, we would not expect \emph{UM2} to show significant variability, or to display the same $\sim3$-day lag as the other far-UV bands.

\subsubsection*{A second, compact reprocessor?}

We see suggestive evidence of a second reprocessing component at near-zero lag, when analyzing the lightcurves after removing the low-frequency variability (\S \ref{sec:results_rm_filtering}). A near-zero lag between the driving continuum and the UV bandpasses is entirely consistent with the `lamp-post' model, given the time sampling of our 2017--2018 monitoring. Thus, this faint second reprocessor may in principle be due to a standard thermal disk. However, most of the reprocessing still occurs at a $\sim3$-day lag in the filtered data. Recently, evidence for multiple UV--optical reprocessing regions has been found for the Type 1 AGN NGC 5548 \citep{Cackett2021}, NGC 4151 \citep{Edelson2017}, and Mrk 279 \citep{Chelouche2019}. The indications of a second reprocessor in Mrk 590 thus add support to the broader emerging picture that AGN UV--optical continuum emission is produced by a multi-component reprocessing geometry. In particular, our data indicate that multiple UV continuum-emitting components may be required even for a recently re-ignited CLAGN.

\subsubsection*{The diffuse BLR continuum contribution to the delays:} 
\defcitealias{Korista2019}{K19}
The broad line region (BLR) produces diffuse continuum emission as it is photoionized by the continuum source \citep[e.g.,][]{Korista2001}. The diffuse continuum responds to changes in the ionizing continuum with a time delay corresponding to the size of the BLR, `diluting' the disk reverberation signal and biasing the lag measurements towards longer values \citep{Korista2001,Lawther2018,Korista2019}. If the X-ray emission is a proxy for the BLR-ionizing continuum, the measured X-ray to UV lags will suffer the same bias. An extended diffuse continuum component might also explain the evidence for two reprocessing regions (\S \ref{sec:results_rm_filtering}). Here, we assess whether diffuse BLR continuum can explain the observed X-ray to UV lags. The reappearance of broad emission lines within one month of the flare-up \citep{Raimundo2019} confirms that broad-line emitting gas was present during 2017. However, no reverberation mapping size estimate for the BLR is available for that epoch. Instead, we use the BLR radius-luminosity relationship \citep[e.g.,][]{Bentz2013} to appropriately scale the diffuse continuum model presented by \citet{Korista2019} (hereafter \citetalias{Korista2019}), allowing us to predict the lag contribution expected due to diffuse continuum.

\citetalias{Korista2019} model the BLR of NGC 5548, which is more luminous than Mrk 590. Assuming that the physical conditions in the BLR scale simply as $r^{-2}$ for a given continuum luminosity, efficiently diffuse continuum-emitting gas will be located at smaller radii in Mrk 590 compared to NGC 5548, with correspondingly shorter diffuse continuum lags.
The appropriate BLR size rescaling depends on the luminosity ratio of the two AGN. It turns out that \emph{the exact luminosity ratio is not important for the following arguments, as long as it does not exceed $\sim1$}; therefore, a crude estimate suffices. The continuum luminosity of the SED used to calculate the \citetalias{Korista2019} BLR model is $\lambda L_\lambda(1138\mathrm{\,\text{\AA}})=3.86\times10^{43}$ erg s$^{-1}$. For Mrk 590, the far-UV flux is highly variable during our monitoring campaign; we adopt $\lambda L_\lambda(1928\mathrm{\,\text{\AA}})=8.20\times10^{42}$ erg s$^{-1}$, i.e., half of the maximum value observed during 2017-2018. As we only need a rough estimate of the luminosity ratio, we neglect the SED slope between 1928 \AA\,\,and \mbox{1138\,\AA}. The size ratio is then related to the luminosity ratio as $R_\mathrm{BLR}\propto L^{1/2}$. We estimate that the BLR in Mrk 590 during 2017 is $\sim$0.45 times as large as that of NGC 5548.

The predicted `as-observed' lags for NGC 5548 as a function of wavelength are presented in Figure 10 of \citetalias{Korista2019}. In principle we would expect the lags induced by diffuse continuum to be shorter for Mrk 590, due to its smaller BLR. The expected lag in \emph{Swift} $UVW2$ ($\sim$1928 \AA) for the \citetalias{Korista2019} model is of order 0.2--0.5 lightdays. Thus, diffuse continuum contamination cannot explain the measured $\sim3$-day lag in $UVW2$. Importantly, given the inherent uncertainties of our BLR size estimate, this result holds even if the BLR in Mrk 590 is in fact as large as that of NGC 5548.

For the \emph{Swift} $U$ band, the \citetalias{Korista2019} model predicts a lag of 1--3 lightdays for NGC 5548. This is longer than the predicted far-UV lags, due to the prominent Balmer continuum produced in the BLR. After rescaling the BLR size, we would expect delays of 0.45--1.35 lightdays in $U$ for Mrk 590 due to diffuse continuum emission. If we consider the \emph{U} band \emph{in isolation}, much of the observed \emph{U} band lag can indeed be attributed to diffuse continuum emission. However, the overall shape of the observed UV--optical lag spectrum (Figure \ref{fig:lamppost}) is inconsistent with this interpretation. Both the far-UV and $U$ lags are identical within their uncertainties, showing no evidence of an excess $U$ band lag due to the Balmer continuum feature.

While some diffuse continuum contamination appears unavoidable for broad-line producing AGN \citep[e.g.,][]{Lawther2018,Korista2019}, we conclude that its contribution to the observed X-ray to UV delays for Mrk 590 is modest. In particular, the observed far-UV lags appear too long to be attributed to diffuse continuum. We note two important caveats for Mrk 590. Firstly, the diffuse continuum contribution to the observed lag depends on the relative luminosities of diffuse continuum and disk emission. The `as-observed' lag spectrum presented by \citetalias{Korista2019} assumes a standard accretion disk for the incident continuum. If the BLR-ionizing continuum is bright for Mrk 590, but the UV-emitting disk is not yet fully formed, the observed lags will resemble those of the diffuse continuum component alone. This would, however, lead to an even more prominent `lag peak' in the $U$ band due to the Balmer continuum. As we do not observe \emph{any} significant lag peak in the $U$ band relative to the far-UV, it is difficult to infer the absolute strength of the diffuse continuum component, but it is likely not the primary driver of the far-UV lags. Secondly, the \citetalias{Korista2019} models are based on the observed continuum variability and emission-line strengths of NGC 5548, a typical Type 1 AGN. The actual variability behavior and BLR physics of a recently reignited CLAGN may affect the diffuse continuum contribution. Comprehensive modeling of the diffuse continuum for Mrk 590 is beyond the scope of this work - but may be an important avenue for future study.

\subsubsection*{A distant X-ray source?}

The most simple `toy model' that can explain the observed $\sim3$-day X-ray to UV delay, with inter-band UV--optical lags consistent with zero, is an X-ray source irradiating the accretion disk at a distance of $\sim3$ light-days. Our initial `lamp-post' modeling (Figure \ref{fig:lamppost}) assumed a compact X-ray source very close to the central black hole. This assumption seems reasonable, given that the X-ray emission is thought to be due to Compton up-scattering of photons produced by the innermost accretion disk \citep[e.g.,][]{Haardt1993,Petrucci2000,Lusso2016}. However, \citet{Kammoun2019} demonstrate that the `lamp-post' model can explain the observed $\sim1$-day X-ray to UV lags for NGC 5548 \emph{only if the X-ray source is located at a height $\sim60r_g$ above the accretion disk}. Here, $r_g$ is the gravitational radius of the central black hole, $r_g=GM_\mathrm{BH}/c^2$. While the full relativistic treatment of the lamp-post model is beyond the scope of this work, the relationship between X-ray source height $h$ and lag becomes almost linear for large $h$ \citep{Kammoun2021}. A rough estimate can be obtained simply by dividing the observed X-ray to UV delay by $2r_g/c$. For Mrk 590, the 3-day X-ray to UV lag would correspond to an X-ray source located $\sim700r_g$ above the accretion disk. It is not immediately obvious how such a distant X-ray corona could be powered, or whether it could generate sufficient luminosity to produce the observed UV variability. We will investigate this scenario in a forthcoming analysis of \emph{XMM-Newton} and \emph{NuSTAR} observations of Mrk 590 (Lawther et al., \emph{in prep.}). \citet{Yang2021} find a marginally-extended radio source that could be a compact, parsec-scale jet in Mrk 590. We speculate that emission from the inner regions of this compact radio jet may potentially provide a $\sim3$-lightday distant X-ray source in Mrk 590. For stellar-mass black holes, increased jet activity occurs during accretion outbursts \citep[e.g.,][]{Fender2009}. If AGN jet production is analogous to that of X-ray binaries, the outburst in Mrk 590 during 2017 might have produced a corresponding increase in power of the inner jet. However, the details of X-ray production in the unresolved inner regions of jets are not well understood. The distant X-ray source scenario also fails to explain the possible presence of a second, compact reprocessor (\S \ref{sec:results_rm_filtering}).

\subsubsection*{Shielding of the outer disk?}

A key assumption of reverberation mapping is that the reverberation timescales are governed by light travel time in the AGN central engine. Given the unexpectedly long X-ray to UV lags (and weak correlations) observed for most AGN with disk reverberation mapping data \citep[e.g.,][]{Shappee2014,Cackett2018,Edelson2019,Cackett2020,Oknyansky2021}, it is worth reconsidering this assumption. Motivated by the $\sim1$-day X-ray to UV lag and the weak correlation strength observed for NGC 5548, \citet{Gardner2017} suggest that a Comptonized inner disk with a large scale height can shield the UV--optical disk from direct X-ray illumination. In this scenario, the delay in UV response to X-ray variability is not primarily due to light travel time, but instead due to the timescale on which the Comptonized inner region expands and contracts due to X-ray heating. The fastest timescale upon which this expansion can occur is the dynamical timescale, $t_\mathrm{dyn}=1/\sqrt{GM_\mathrm{BH}/r^3}$. For Mrk 590, the dynamical timescale is $\sim3.5$ days at a radius of 70$r_g$; \citet{Gardner2017} argue that the disk is likely Comptonized at smaller radii for AGN accretion disks. Thus, our observed $\sim3$-day X-ray to UV lags are consistent with the shortest expected response time for X-ray heating of a Comptonized inner disk. Assuming this scenario, it is tempting to identify the putative second reprocessor (\S \ref{sec:results_rm_filtering}) with a weak direct reverberation signal as some fraction of the X-rays directly irradiate the inner disk, while the dominant $\sim3-$day response occurs as the Comptonized region changes its size. 

However, we observe that the X-ray and far-UV lightcurves are highly correlated for Mrk 590 (\S \ref{sec:discussion_correlation}). In contrast, for the \citet{Gardner2017} scenario, the UV response to high-frequency X-ray variability is expected to be suppressed, due to the slow response of the Comptonized shielding component. Detailed modeling of this shielding component is required in order to investigate whether the observed UV response for Mrk 590 can in fact be reproduced.

\subsubsection*{The energetics of X-ray reprocessing:} 

In this work, we have neglected the energetics of the reprocessing scenario. If the X-ray source indeed represents the  driving continuum (and not simply a proxy of it), then the question remains whether or not the X-ray flares are sufficiently powerful that their irradiation of the disk can produce the observed response. For Mrk 590 we find very strong evidence that most of the UV variability occurs as a response to a driving continuum that is observed in X-rays $\sim3$ days earlier than in UV. However, this does not conclusively demonstrate that the X-rays \emph{are} the driving continuum. As a counter-example, for the unobscured AGN Ark 120, \citet{Mahmoud2022} demonstrate that reprocessing of a compact X-ray continuum at scale-height $h=10r_\mathrm{g}$ fails to produce enough UV variability to explain their observations. In their modeling, increasing the X-ray corona height to $h=100r_\mathrm{g}$ increases the resulting UV variability, but produces a stronger than observed X-ray to UV correlation (which is rather weak for Ark 120). They suggest that the variability is intrinsic to the UV-emitting region - which cannot be the case for Mrk 590, as the UV clearly responds to the X-rays. In fact, the behavior of Mrk 590 is quantitatively similar to the $h=100r_\mathrm{g}$ model presented by \citet{Mahmoud2022}. Mrk 590 is unusually X-ray bright for its UV luminosity \emph{(Lawther et al., in prep)}, which may also explain the observed strong UV response. Energetically consistent modeling would likely help constrain the scenarios already outlined in this Discussion (\emph{e.g.,} secondary reprocessing regions and/or distant X-ray sources). We will address these issues in future work, as part of a full analysis of the observed spectral energy distribution and its variability.

\section{Conclusion}\label{sec:conclusion}

The changing-look AGN Mrk 590 lost its UV continuum and broad-line emission components at some point between 2006 and 2013 \citep{Denney2014}. The X-ray and UV emission brightened in August 2017, and has displayed repeated flare-ups since then. It has not yet returned to the historic low-luminosity state observed in 2013, nor to its historic high-luminosity state of the late 1980s, when it was a \emph{bona fide} Type 1 AGN. As the broad emission lines reappeared during the initial flare-up \citep{Raimundo2019}, the extreme-UV ionizing continuum must also have reappeared; we interpret the UV flares since 2017 as largely due to a highly variable AGN continuum source. The characteristic timescales of the flares, that we determine to be $\sim100$ days, are consistent with the thermal timescales expected in the inner disk, as also observed for non-CLAGN \citep[e.g.,][]{Burke2021}.

Even though Mrk 590 is accreting at a rate that only reaches a maximum of $\sim$5\% of the Eddington rate, its excess variance $F_\mathrm{var}$ during the post-2017 flaring is comparable to those of AGN accreting near the Eddington limit. The X-ray to UV variability correlation for Mrk 590 is among the strongest observed for AGN monitored by \emph{Swift}. Thus, it appears that intense X-ray outbursts directly illuminate the UV-emitting region and produce an unusually strong UV response. As non-CLAGN often display weak X-ray to UV correlations \citep[e.g.,][]{Edelson2019}, we speculate that recently re-ignited CLAGN may differ from steady-state sources in terms of the relative geometries of the X-ray and UV-emitting regions, and/or in terms of the obscuration of the X-ray emitter as seen from the disk. Two other CLAGN demonstrate similar strong X-ray to UV correlations immediately after their `flare-up' phases \citep{Oknyansky2017,Oknyansky2021}. More high-cadence monitoring observations of CLAGN are required to confirm this trend.

We constrain the UV--optical inter-band lags to be less than $\sim1.5$ days, which is consistent with the expectations for reverberation in a thermal disk. Surprisingly, the UV flares lag the X-rays by $\sim$3 rest-frame days. This long lag is inconsistent with the standard `lamp-post' model for X-ray irradiation driving UV variability. Simply increasing the size of a standard thermal disk model cannot reproduce the observed lag spectrum. While a $\sim3$-day X-ray to UV delay is suggestive of reprocessing in the inner BLR, the lack of a Balmer continuum feature in the observed lag spectrum likely rules out a major contribution from diffuse BLR continuum emission. Other possibilities include \emph{a)} an X-ray source at a large scale-height above the disk \citep[e.g.,][]{Kammoun2019}, \emph{b)} an additional UV reprocessor located $\sim3$ light-days form the central source, or \emph{c)} that the lag is not due to light travel time. In the latter case, the Comptonized inner disk shielding mechanism presented by \citet{Gardner2017} could produce the $\sim3$-day lag, but would also suppress the response to high-frequency X-ray variability. This is difficult to reconcile with the observed strong X-ray to UV correlation. 

We find suggestive evidence of a second UV reprocessor at near-zero lag. The second reprocessing component is fainter than the dominant $\sim3$-day lagged component, and is only revealed after filtering out slow variability in the X-ray and UV lightcurves. Intriguingly, the standard `lamp-post' reprocessing model predicts a near-zero X-ray to UV lag for Mrk 590. In scenarios where the UV response is dominated by distant reprocessing regions, this faint component may thus represent the response from the accretion disk itself, as suggested by \citet{Cackett2021} for NGC 5548. Alternatively, for the scenario proposed by \citet{Gardner2017}, we speculate that the Comptonized inner disk may produce a faint, prompt UV response to direct X-ray illumination.

Further insights into the accretion physics during these flares, and during the previous low-luminosity state, can be gleaned through \emph{1)} studying the evolution of the spectral energy distributions before and during the flares; \emph{2)} determining the evolution of the soft X-ray excess, and its dependence on continuum luminosity, using deep X-ray and UV observations; \emph{3)} performing energetically consistent simulations to explore various reprocessing scenarios; and \emph{4)} testing the periodicity of the post-2017 flaring activity. We will address these issues in future work. 

\paragraph*{Acknowledgements:} We thank the anonymous referee for their insightful comments and suggestions, which both improve the quality of the present work and provide inspiration for future investigations. Dr. Ying Zu provided valuable guidance regarding our interpretation of the best-fit top-hat widths in our \textsc{Javelin} analysis. Much of the analysis presented in this paper relies on the \textsc{HEASoft}, \textsc{Ftools}, and \textsc{XSPEC} software packages and online resources. This work was supported by the Indepedent Research Fund Denmark via grants DFF-4002-00275 and DFF-8021-00130. DL acknowledges financial support from NASA through award numbers 80NSSC20K1484 and 80NSSC20K0635.

\paragraph*{Data availability:} The flux measurements used in the lightcurves are available in the article and in its online supplementary material. The raw observational data is available from the NASA/GSFC public \emph{Swift} archive, e.g., \url{https://heasarc.gsfc.nasa.gov/cgi-bin/W3Browse/swift.pl} .

\bibliographystyle{mnras}
\bibliography{bibliography}

\appendix

\section{Structure function analyses}\label{sec:appendix_SF}

\begin{figure*}
     \newcommand{\picscale}{0.52}
     \newcommand{\pdfscale}{0.30}
     \centering
     \includegraphics[scale=\picscale,trim={0 0 0 0}, clip]{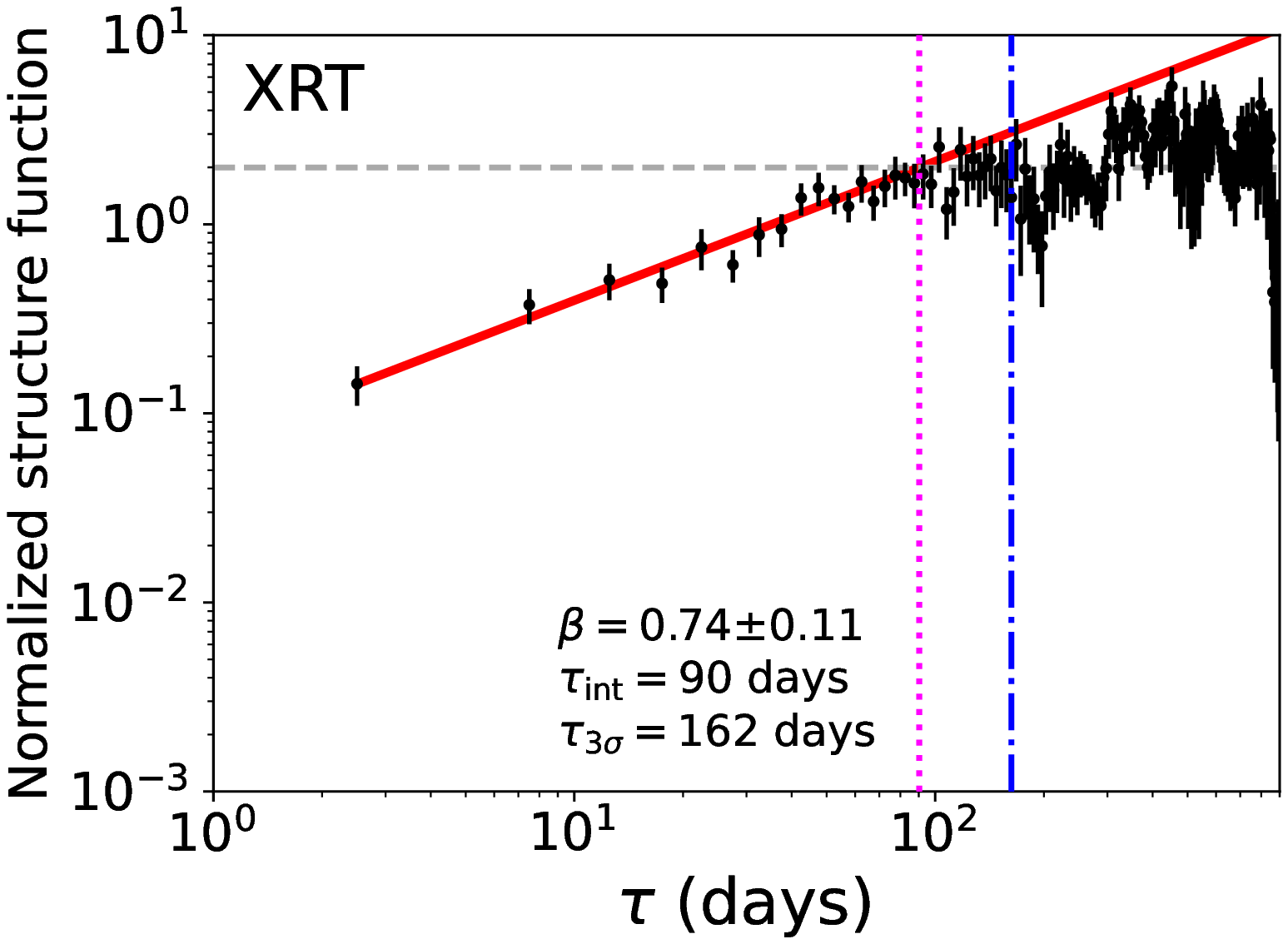}
     \includegraphics[scale=\picscale,trim={0 0 0 0}, clip]{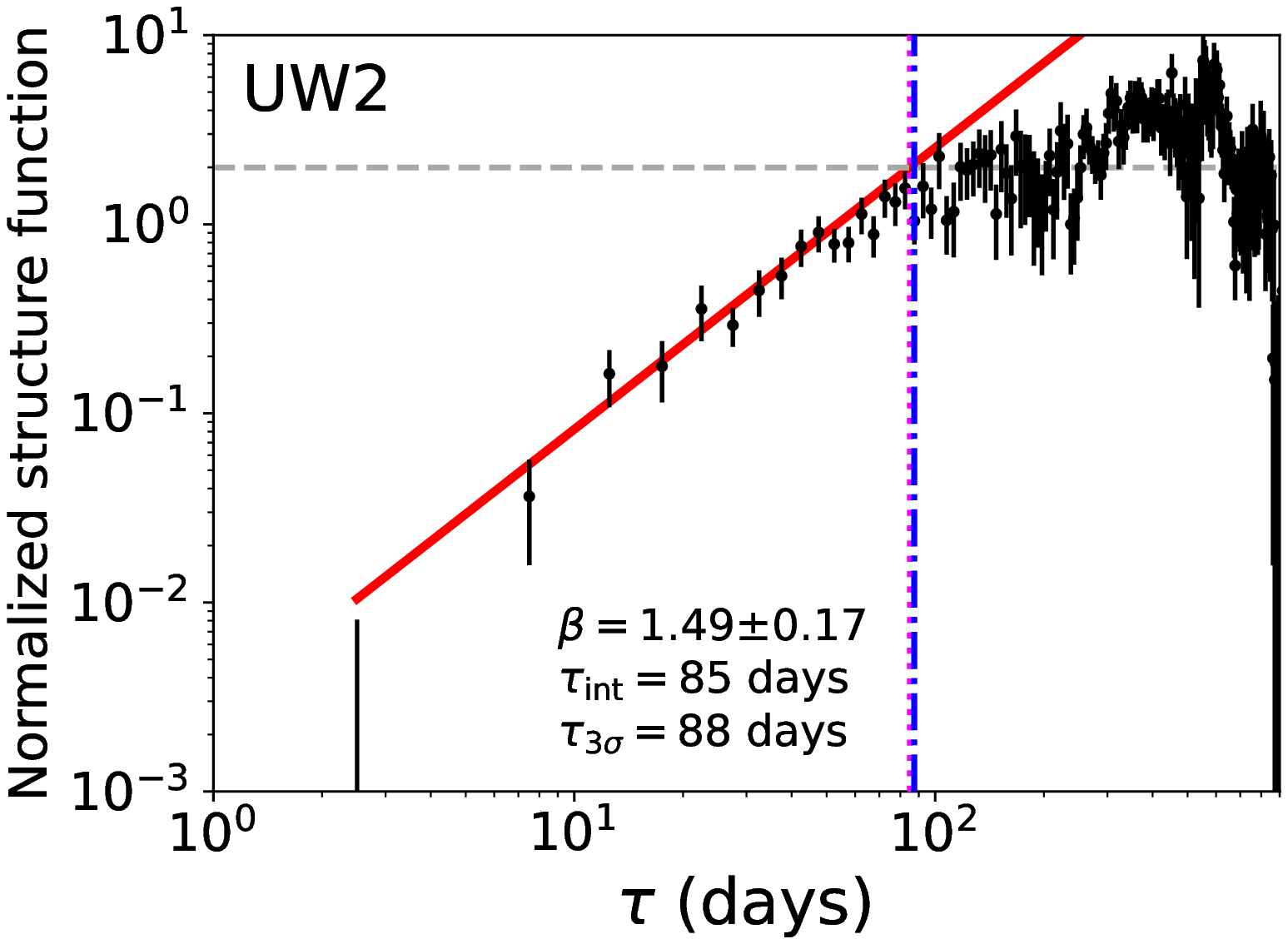}
     \includegraphics[scale=\pdfscale,trim={0 0 0 0}, clip]{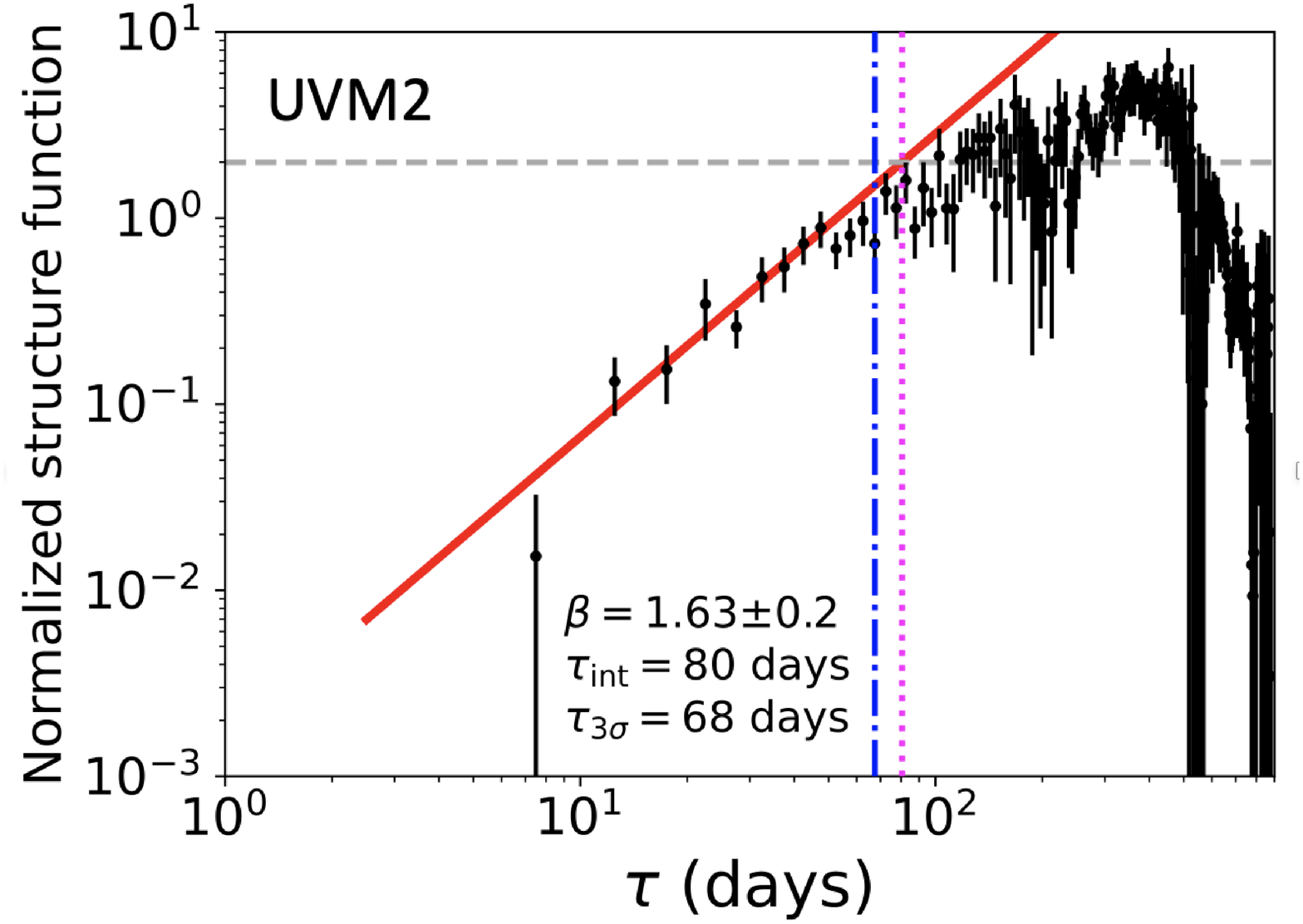}
     \includegraphics[scale=\pdfscale,trim={0 0 0 0}, clip]{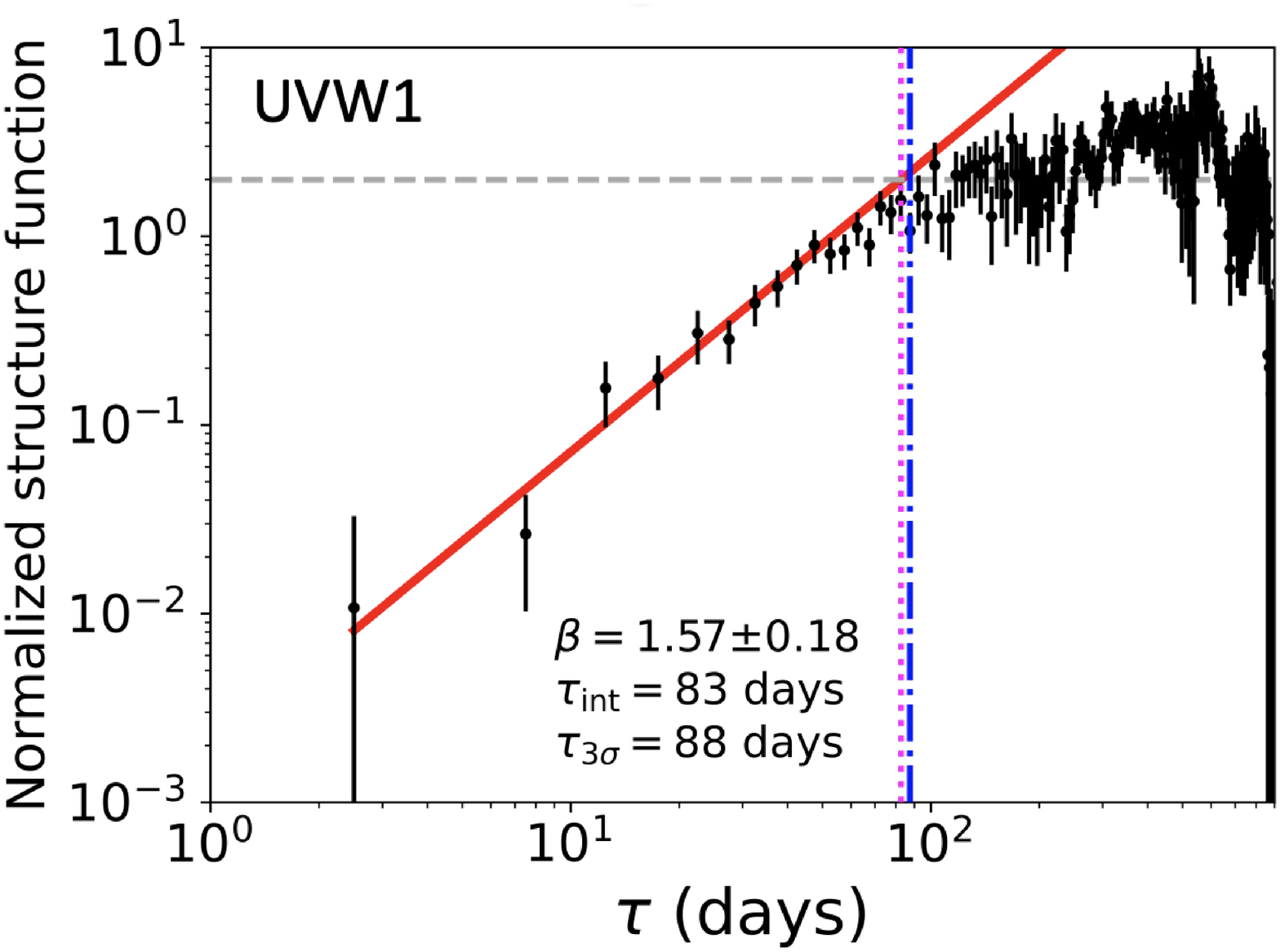}
     \includegraphics[scale=\picscale,trim={0 0 0 0}, clip]{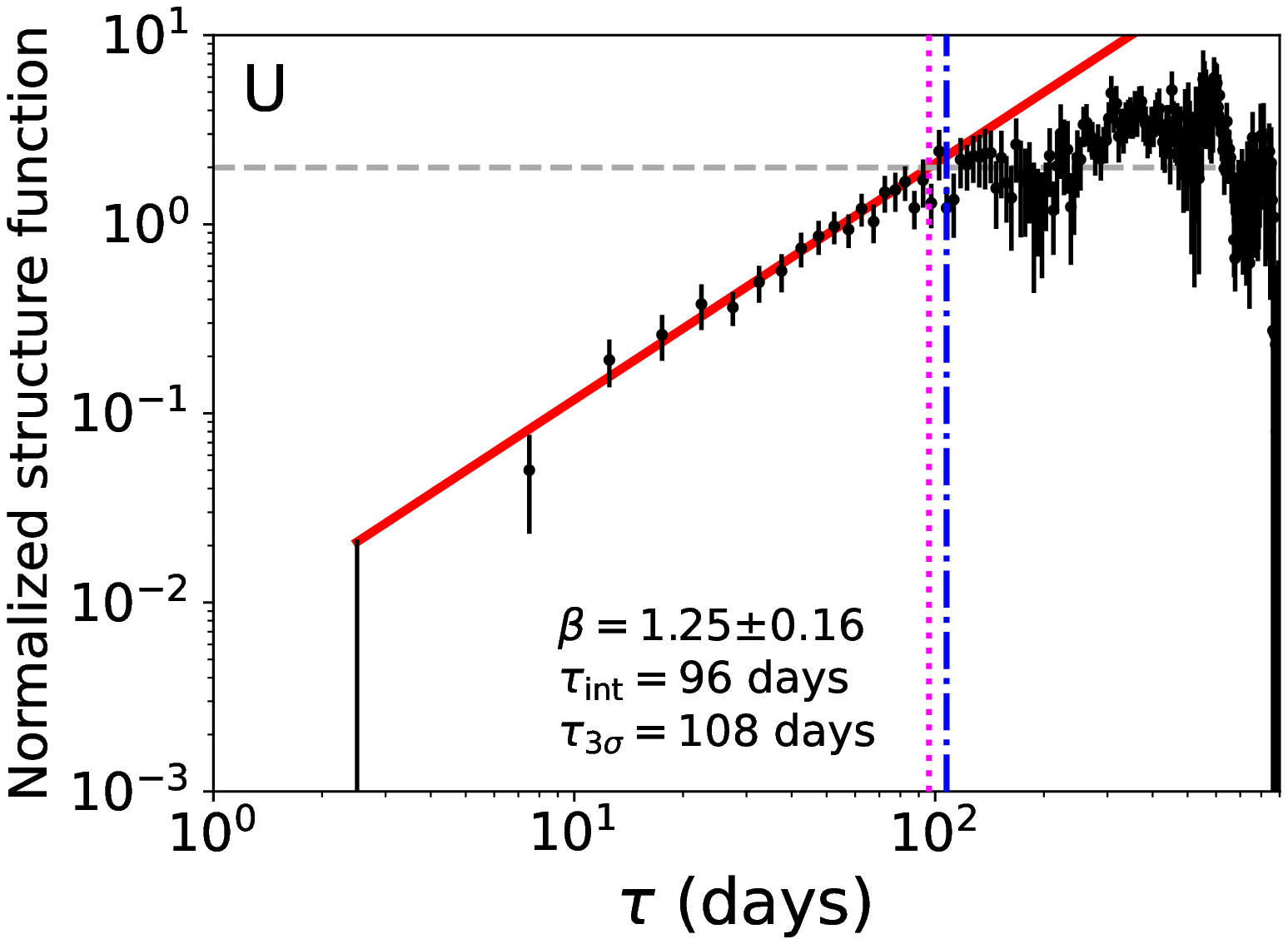}
     \includegraphics[scale=\picscale,trim={0 0 0 0}, clip]{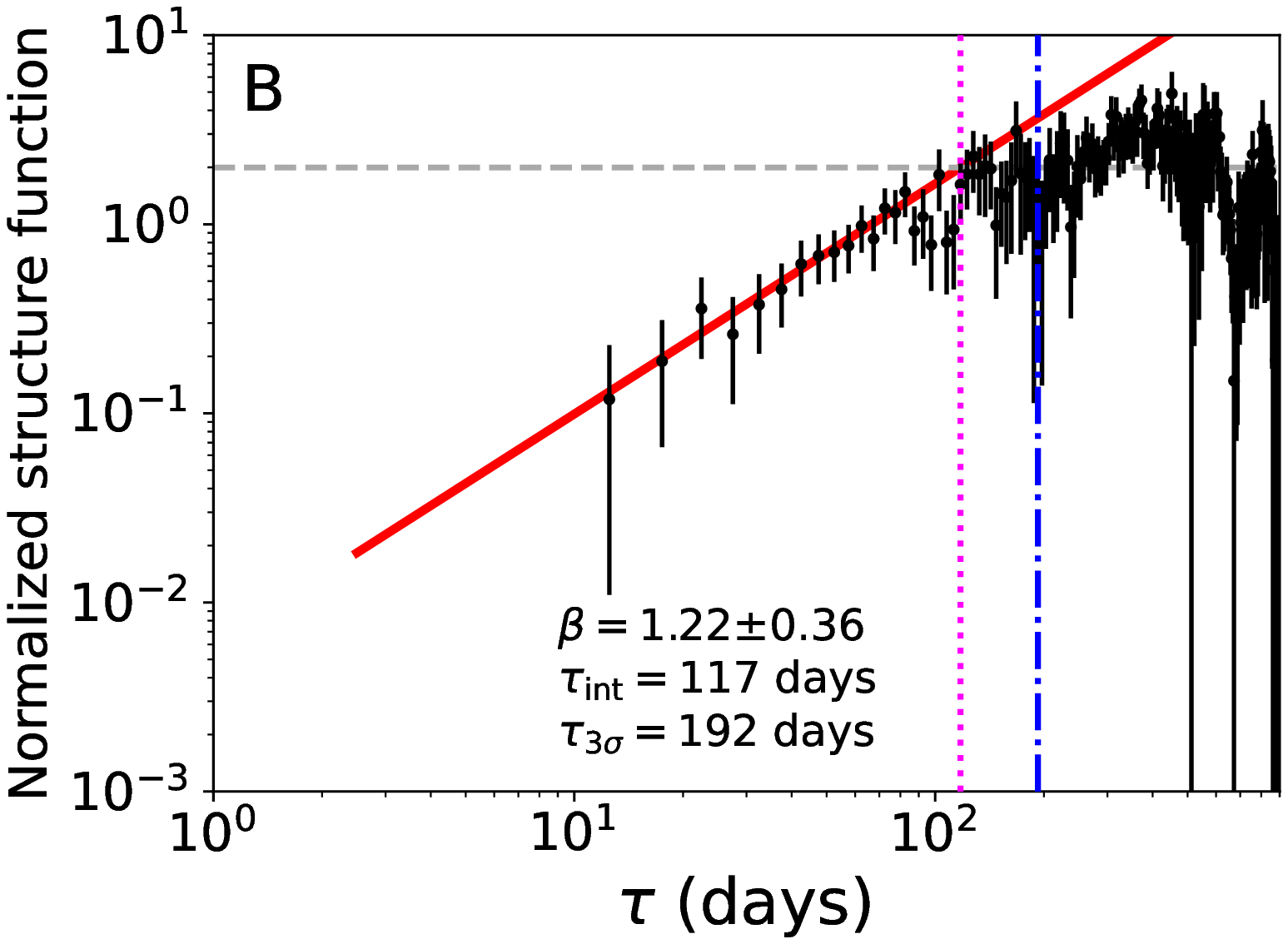}
     \caption{Binned structure functions for the \emph{Swift} XRT and UVOT lightcurves since August 2017 (\emph{black circles and error-bars}), normalized to the observed variance $\sigma^2$ of each lightcurve (\emph{gray dashed line}). We model the structure functions as a power-law at timescales $5<\tau<50$ observed-frame days, and extrapolate this model to longer delays (\emph{solid red line}). The best-fit power-law slope $\beta$ is most shallow for XRT, indicating more variability on short timescales in the X-rays than in the UV--optical. We see no obvious deviation from the power-law before the structure function reaches the variance level. We make two rough estimates of the associated characteristic timescale, as follows. The intersection $\tau_\mathrm{int}$ between the extrapolated power-law model and the lightcurve variance ranges between 80 and 117 observed-frame days (\emph{magenta dotted lines}). The timescale $\tau_{3\sigma}$ is the shortest timescale at which the observed structure function deviates from the power-law model by more than three times its uncertainty (\emph{blue dot-dash lines}). These estimates are in rough agreement except for the XRT and \emph{B}-band lightcurves, for which $\tau_{3\sigma}$ is somewhat longer. The total monitoring duration for these lightcurves is 1337 observed-frame days, i.e., a factor $\sim10$ longer than these estimated characteristic timescales. The structure function for the \emph{V} band is very noisy; we do not show it here.}
     \label{fig:structure_function}
\end{figure*}

Here, we describe the structure function analysis summarized in \S \ref{sec:results_lc} in detail. Following \citet{Collier2001ApJ}, we calculate the binned first-order structure function for each lightcurve as

\begin{equation*}
    SF(\tau)=\frac{1}{N(\tau)\sigma_\mathrm{lc}^2}\left(\sum_{i<j}\left[ f(t_i)-f(t_j)\right]^2\right)-2\delta^2,
\end{equation*}

\noindent where $N(\tau)$ is the number of flux measurements $f(t_i)$, $f(t_j)$ for which the time delay $\tau=t_j-t_i$ is contained in a given $\tau$ bin. We subtract two times the variance of the mean photometric error, $2\delta^2$, to remove the variability expected due to measurement errors in each bin, and normalize the structure functions by the lightcurve flux variance $\sigma_\mathrm{lc}^2$. We choose a bin size of five observed-frame days. This ensures at least six pairs of observations in each bin, which provides a reasonable time resolution (given the observational cadence and timescales involved) and sufficiently high signal-to-noise in $SF(\tau)$ for our modeling purposes. The median number of pairs per $\tau$-bin is 34 for \emph{UM2} and $\approx50$ for the other lightcurves. We note that the structure function for time delays $\tau\lesssim14$ observed-frame days is largely determined by the observations during 2017--2018, for which the average time sampling is $\sim2$ days, yielding the majority of the observation pairs at short time separations. We present the structure functions for the 0.3--10 keV X-ray band, and the \emph{UVW2} through \emph{B} filters, in Figure \ref{fig:structure_function}. The structure functions for the soft or hard X-ray bands are near-identical to the 0.3--10 keV band, as expected given the low X-ray spectral variability. The UVOT \emph{V} structure function is very noisy, and not suitable for analysis, due to the low \emph{V}-band excess variance.

For purely stochastic variability, the noise-subtracted structure function will reach a plateau of $2\sigma_\mathrm{lc}^2$ at long timescales, and will follow a power-law at shorter timescales for which the variability is correlated \citep{Collier2001ApJ}. We model the structure functions as a power-law at timescales $5<\tau<50$ observed-frame days, and extrapolate to longer timescales (Figure \ref{fig:structure_function}, solid red lines). Given the large uncertainties on the power-law slope, our measured $\beta=1.22\pm0.36$ in the \emph{B} band is consistent with the optical monitoring of Mrk 590 during its high-luminosity state in 1989, for which $\beta=0.84\pm0.20$ \citep{Collier2001}. The power-law slope $\beta$ is more shallow in the X-rays ($\beta=0.74\pm0.11$) than in the far-UV bands (e.g., $\beta=1.49\pm0.17$ for \emph{UVW2}). \citet{Gallo2018} also note a difference in power-law slopes between X-ray and UV--optical lightcurves for their \emph{Swift} monitoring of Mrk 335. These authors find a very flat X-ray structure function with $\beta=0.14\pm0.05$, and attribute this to the X-ray variability having a much shorter characteristic timescale. For Mrk 590, the correlation between X-ray and UV lightcurves is very strong (\S \ref{sec:results_rm}), and the `flattening' of the X-ray structure function at timescales of $\sim100$ days is well-determined. As discussed in \S \ref{sec:results_lc}, we find evidence for a rapid variability component in our \textsc{Javelin} modeling of the X-ray lightcurve. We can thus attribute the flatter X-ray structure function to the presence of this rapidly varying component. We speculate that its absence in the UV lightcurves may be due to a `smearing' of the most rapid X-ray variability in an extended reprocessor, which would suppress the short-timescale variability in the reprocessed lightcurves.

We make two rough estimates of the characteristic timescale at which the structure function flattens, as follows. The intersection $\tau_\mathrm{int}$ between the extrapolated power-law model and the `plateau' at $2\sigma_\mathrm{lc}^2$ corresponds to the expected flattening for a process governed by a single characteristic timescale. We derive uncertainties on $\tau_\mathrm{int}$ from the $1\sigma$ uncertainties on the power-law amplitude and exponent. We find $80<\tau_\mathrm{int}<117$ observed-frame days for the XRT and UVOT lightcurves (excluding \emph{V}). We also determine the shortest timescale for which the binned structure function deviates from the power-law model by more than three times its uncertainty, and denote this timescale $\tau_{3\sigma}$. These two estimates are broadly consistent, except for the XRT and the UVOT \emph{B} bands, for which $\tau_{3\sigma}$ is longer. Rarely, AGN display an `early flattening' in their structure functions, suggesting variability on two characteristic timescales \citep[e.g.,][]{Collier2001ApJ,Gallo2018}. For Mrk 590, this does not appear to be the case, as $\tau_\mathrm{int}\lesssim\tau_{3\sigma}$ for all our lightcurves. For this reason, and as $\tau_{3\sigma}$ is very sensitive to the statistical uncertainty in each bin, we only discuss $\tau_\mathrm{int}$ elsewhere in this work.

\citet{Collier2001ApJ} demonstrate that the measured timescales for which $SF(\tau)$ flattens can be underestimated (relative to the `true' characteristic variability timescale of the source) if they exceed approximately one-third of the total monitoring duration. As we measure characteristic timescales of $\sim100$ observed-frame days, for a monitoring duration of 1337 days since August 2017, our structure functions are not affected by this bias. This allows us to more confidently assign physical significance to the measured $\tau_\mathrm{char}$ estimates in our discussion of the flaring behavior (\S \ref{sec:discussion_flares}).

\section{Biases due to time sampling in our reverberation mapping analysis}\label{sec:appendixA}

\defcitealias{Edelson2019}{E19}

\begin{figure*}
    \includegraphics[scale=0.53]{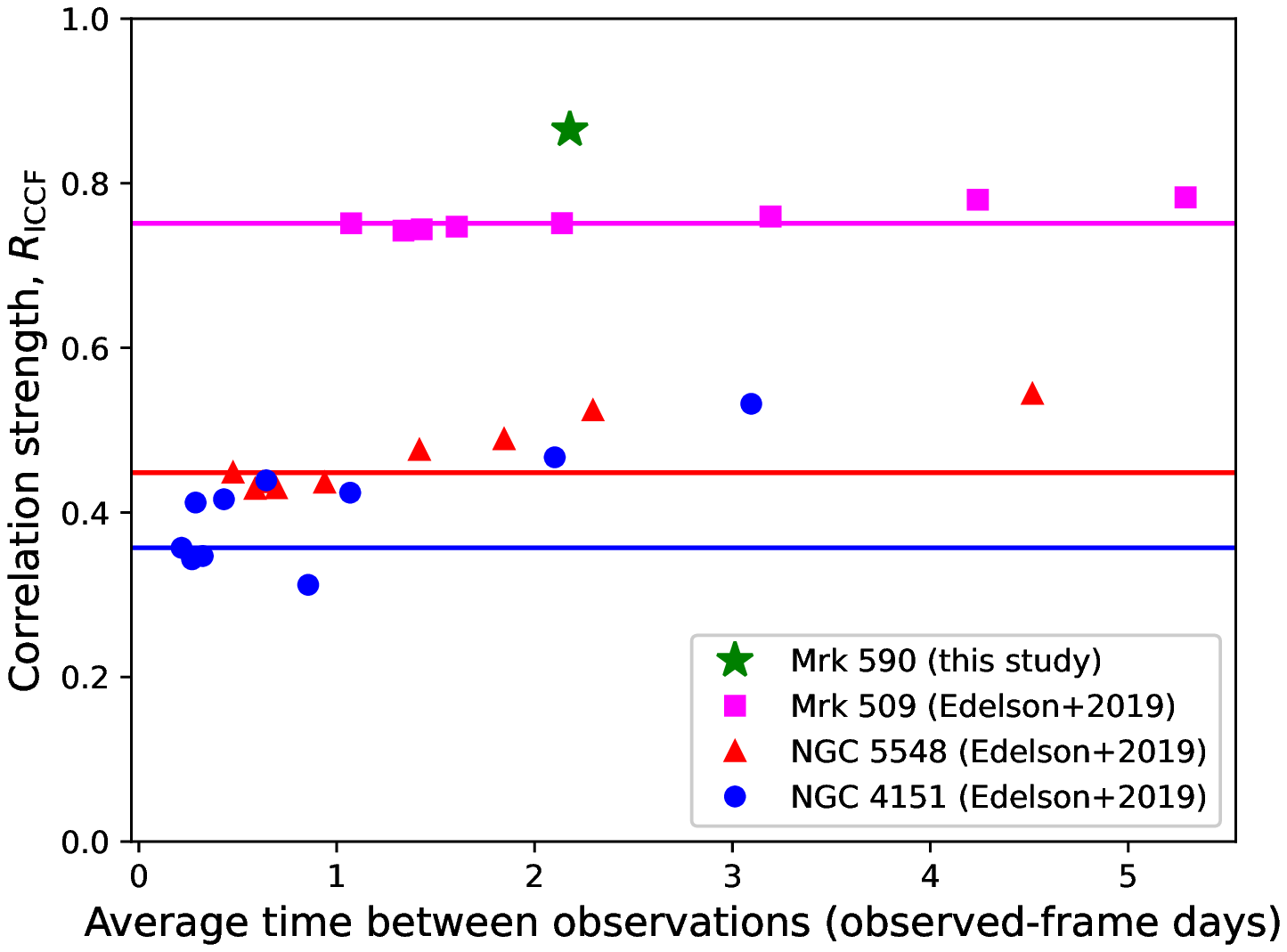}
    \includegraphics[scale=0.53]{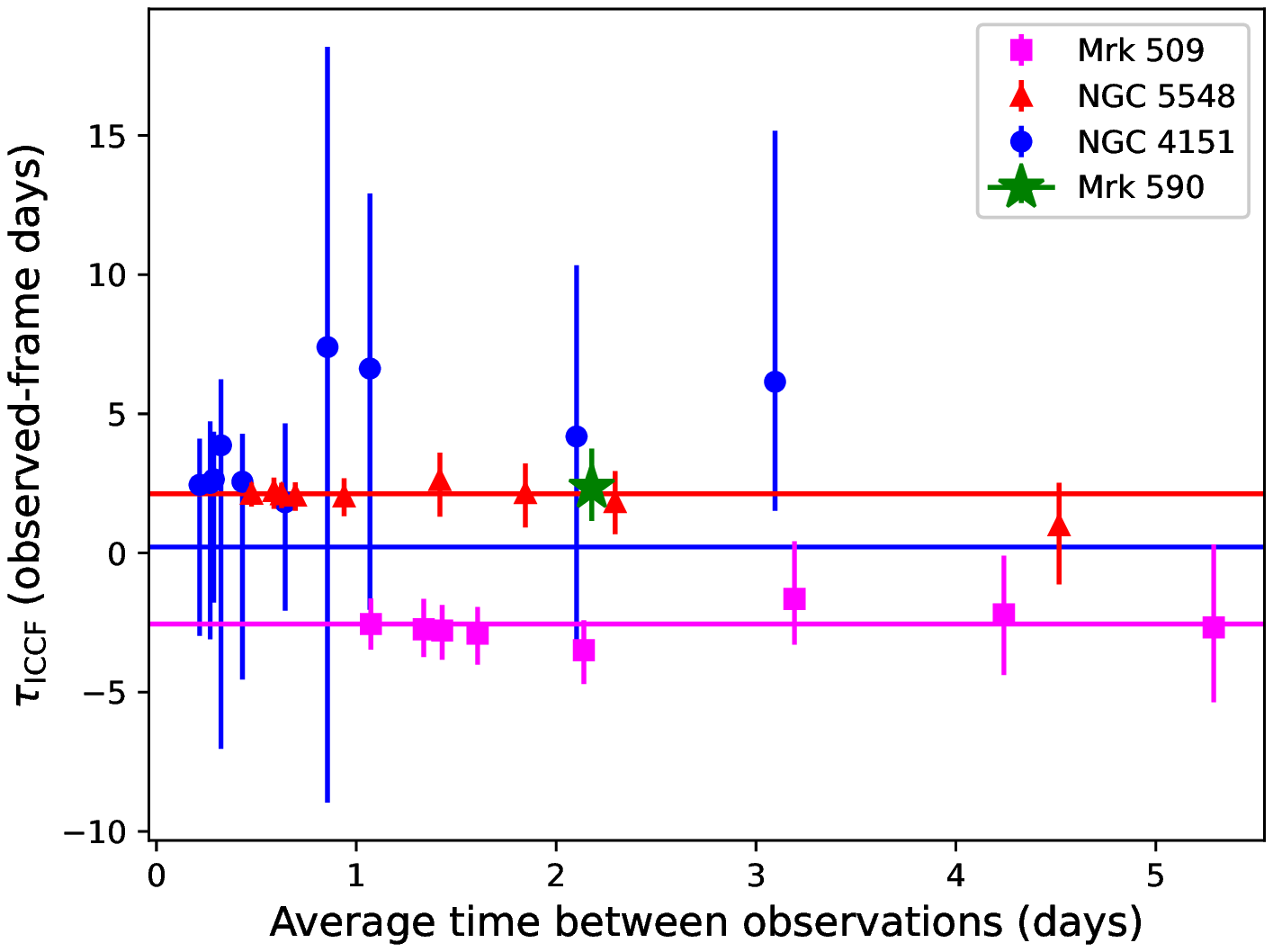}
    \caption{\emph{Left:} Measured $R_\mathrm{ICCF}$ as a function of average time separation between observations, for the UVOT $UVW2$ lightcurves (relative to XRT) presented by \citet{Edelson2019}. We artificially degrade these lightcurves by removing every $n$th data point, producing a range of time separations for each lightcurve, in order to investigate the influence of observational cadence. Horizontal lines represent the `true' $R_\mathrm{ICCF}$ for the observed lightcurves. For comparison, we find $R_\mathrm{ICCF}=0.87$ for Mrk 590 in 2017--2018, with an average observational cadence of $\sim2$ days (green star). \emph{Right:} Measured time delay $\tau_\mathrm{ICCF}$ versus average time separation, for the artificially degraded lightcurves. Horizontal lines represent the `true' $\tau_\mathrm{ICCF}$ values.}
    \label{fig:time_sampling}
\end{figure*}

We measure an X-ray to $UVW2$ correlation strength of $R_\mathrm{ICCF}=0.87$ for Mrk 590 during 2017--2018, with similar strong correlations between the X-rays and other UV bands (\S \ref{sec:results_rm}). To our knowledge, this is the strongest X-ray to UV correlation yet observed for AGN with available intensive reverberation mapping data, which tend to display stronger inter-band UV--optical correlations than X-ray--UV (\S \ref{sec:discussion_correlation}). However, dedicated disk reverberation mapping campaigns with \emph{Swift} usually have better time sampling (observing every $\sim$0.5--1 days) than do our 2017--2018 data (on average 2.2 days). To allow a robust comparison between our results and previous studies, it is important to determine whether our measurement of $R_\mathrm{ICCF}$ tends towards higher values due to our more sparse sampling. Such a bias would occur, e.g., if fast X-ray variability on timescales of $<2$ days produces a weaker UV response than slower variability. 

\citet{Edelson2019} (hereafter, \citetalias{Edelson2019}) present continuum reverberation mapping studies using \emph{Swift} for four AGN: NGC 5548, Mrk 509, NGC 4151, and NGC 4593. We base the following tests on their published \emph{Swift} data. We note that NGC 4593 ($M_\mathrm{BH}\approx7.6\times10^6M_\odot$) has a lower black hole mass than Mrk 590 ($M_\mathrm{BH}\approx3.7\times10^7M_\odot$), and its accretion disk size is expected to be correspondingly smaller. In fact, \citetalias{Edelson2019} find X-ray to UV time delays $\tau_\mathrm{ICCF}\sim0.6$ days for NGC 4593, which are not detectable when the sampling is degraded to $\sim2$ days. We exclude this object, and perform these tests for the three remaining AGN in the \citetalias{Edelson2019} sample. 

To investigate the bias due to time sampling, we construct several artificially degraded lightcurves for each observed \citetalias{Edelson2019} lightcurve, by discarding individual data points in a uniform manner. We then analyze the X-ray to $UVW1$ correlations for these degraded lightcurves. For NGC 5548 ($R_\mathrm{ICCF}=0.44$) and NGC 4151 ($R_\mathrm{ICCF}=0.36$), where the X-ray to UV correlation in the observed lightcurves is relatively weak, the measured correlation increases as the time sampling becomes more sparse (Figure \ref{fig:time_sampling}, left panel). Thus, there is indeed some bias due to observational sampling. However, the increase is only of order $\Delta R_\mathrm{ICCF}\sim0.1$ when the time sampling is degraded from $\sim0.5$ to $\sim2$ days. For Mrk 509, which displays the highest XRT to $UVW2$ correlation strength of the three AGN presented by \citetalias{Edelson2019}, the effect of degrading the time sampling is minimal. For this AGN, the X-rays lag the far-UV bands, so it is not fully analogous to the behavior we see for Mrk 590. 

We also test whether the measured time delays are biased by degrading the time sampling. Our simulation results do not show evidence of such bias; the  scatter in the ICCF centroid distribution increases as the time sampling is degraded to $\sim2$ days, but with no obvious trends (Figure \ref{fig:time_sampling}, right panel). 

We conclude that our measured correlation strengths $R_\mathrm{ICCF}$ and time delays $\tau_\mathrm{ICCF}$ are rather robust to time sampling issues. While reducing the sampling cadence can bias the measured correlation strength slightly, we would expect to measure $R_\mathrm{ICCF}\gtrapprox0.75$ for Mrk 590 for a $\sim0.5$-day sampling, assuming similar variability in the underlying driving continuum as for the \citetalias{Edelson2019} sample. We see no evidence that $\sim3$-day X-ray to UV lag measurements are biased towards smaller or larger values as the time sampling is degraded to $\sim2$ days. The rather large uncertainties on the measured lags are, however, to some extent due to the sparse time sampling. 

\section{Simulations of boxcar-filtered lightcurves}\label{sec:appendixB}

\begin{figure*}
   \newcommand{\picscale}{0.53}
   \centering
   \includegraphics[scale=\picscale,trim=0 30 0 0, clip]{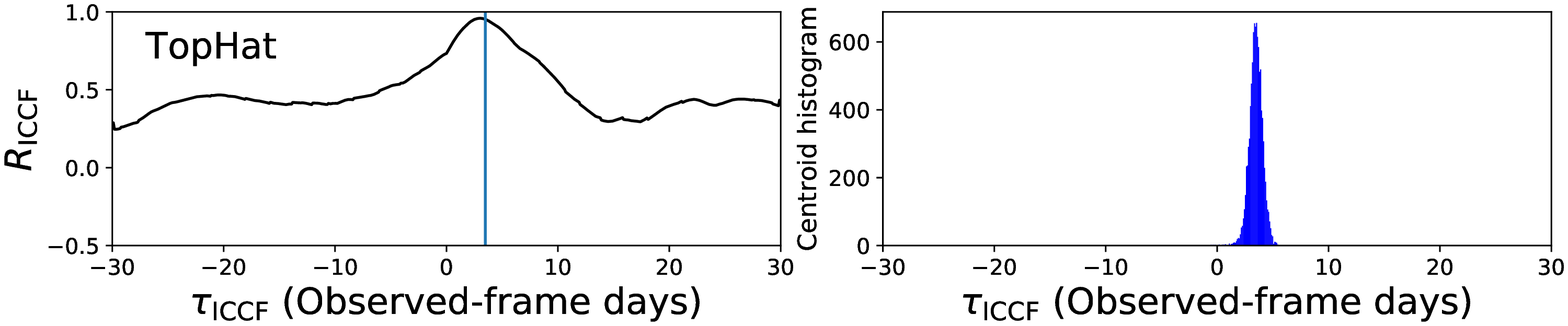}
   \includegraphics[scale=\picscale,trim=0 30 0 0, clip]{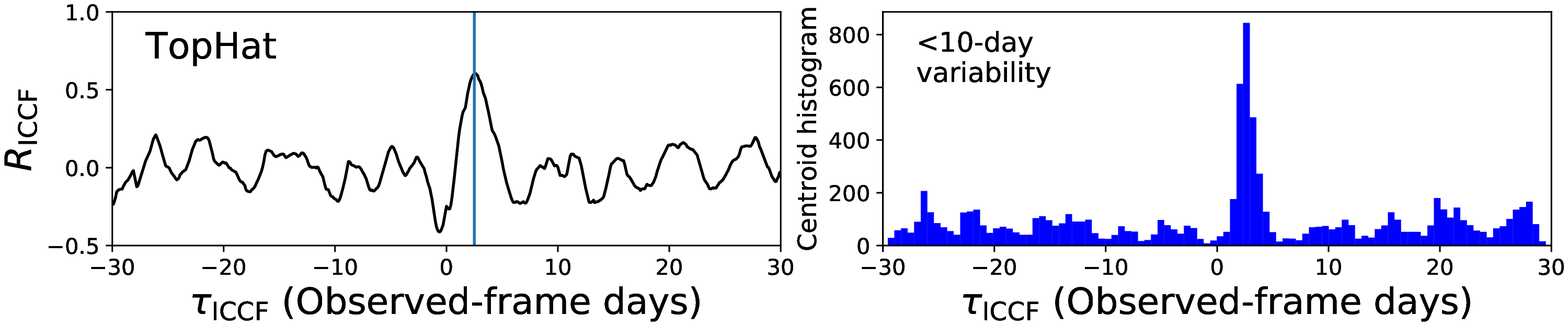}
   \caption{Interpolated Cross-Correlation Function (ICCF) analysis of simulated lightcurves based on the observed 2017--2018 X-ray continuum, as described in Appendix \ref{sec:appendixB}. We simulate a reprocessed lightcurve as a top-hat function at a time delay of 3 light-days and a width of 0.33 light-days. The cross-correlation functions (\emph{left panels}) and centroid distributions (\emph{right panels}) are for the reprocessed lightcurve relative to the model driving continuum. For the unfiltered lightcurves (\emph{top panels}), we see peaks in both the CCF and the centroid distribution at $\sim3$ lightdays. For the filtered lightcurves (\emph{bottom panels}), the overall correlation strength is reduced by the filtering. We note a slight shift of both the CCF and centroid distribution peaks towards shorter delays, but do not see a spurious second peak appear in either distribution.}
   \label{fig:filtering_simulations}
\end{figure*}

In \S \ref{sec:results_rm_filtering} we apply a filtering procedure to our 2017--2018 X-ray and UV lightcurves, and repeat our reverberation mapping analysis. We find that, when applying a 10-day boxcar smoothing width, a second peak at near-zero lag appears in the X-ray to $UVW2$ cross-correlation function. Here, we investigate whether this secondary component is a consequence of the filtering process itself. To this end, we simulate a single reprocessor with no intrinsic response at zero lag, and apply our filtering scheme to the simulated lightcurves, to check whether a spurious zero-lag response appears.

To best match the real observational conditions for our 2017--2018 data, we use the observed X-ray lightcurve during the 2017 flare as the driving continuum for this simulation. We use a quadratic interpolation scheme to obtain a model driving continuum with fine time sampling. We model the reprocessing region as a top-hat response function with a width of $\Delta r$=0.33 light-days and a central delay of $r=3$ light-days. We then resample both the model X-ray and the reprocessed lightcurves to match the time sampling of our observations (i.e., matching the individual timestamps of each observation, not just the average sapling rate). We add Gaussian noise at the 5\% level to each data point, to roughly match the relative uncertainties of our individual UVOT observations. This results in model X-ray and reprocessed lightcurves with similar characteristics to our observed data. To identify the effects purely due to the filtering procedure, we first perform an ICCF analysis on the simulated, \emph{unfiltered} lightcurves (Figure \ref{fig:filtering_simulations}, top panels).

We apply the boxcar-smoothing filtering procedure to both the driving and reprocessed lightcurves, selecting a boxcar width of 10 days, to match the smallest smoothing width applied in \S \ref{sec:results_rm_filtering}. We repeat the ICCF analysis on the simulated, filtered lightcurves. We find that the overall correlation strength is reduced by the filtering procedure (Figure \ref{fig:filtering_simulations}, left panels), as also seen for our real filtered data. We also note a slight ($<1$-day) shift of both the CCF and the centroid distribution peaks towards shorter lags in the filtered data (Figure \ref{fig:filtering_simulations}, bottom panels). Importantly, for these simulated reprocessed lightcurves that contain no \emph{intrinsic} zero-delay component, secondary peaks in the CCF and centroid distributions near zero delay do not appear after applying the 10-day filtering. Thus, the appearance of a secondary CCF peak in our observed data after filtering (\S \ref{sec:results_rm_filtering}) is likely a real feature of the data, and not an artifact of our filtering procedure.

\end{document}